\documentclass[useAMS,usenatbib]{mnras}
\usepackage[english]{babel}
\usepackage[fleqn]{amsmath}
\usepackage{amsfonts}
\usepackage{txfonts}
\usepackage{graphicx}
\usepackage{amssymb}
\usepackage[T1]{fontenc}
\usepackage{ae,aecompl}
\usepackage{fancyhdr}
\usepackage{multicol}
\usepackage{layout}
\usepackage{graphicx}
\usepackage{multirow}
\usepackage{color}
\usepackage{multirow}
\usepackage{times}
\usepackage[outdir=./]{epstopdf}

\newif\ifAMStwofonts
\AMStwofontstrue

\title[Spherical collapse with tidal shear]{Spherical collapse of dark matter haloes in tidal gravitational fields}
\author[Reischke et al.]{Robert Reischke$^{1}$\thanks{e-mail:
reischke@stud.uni-heidelberg.de}, Francesco Pace$^2$, Sven Meyer$^3$ and Bj\"orn Malte Sch\"afer$^{1}$\\
$^{1}$ Zentrum f{\"u}r Astronomie der Universit{\"a}t Heidelberg, Astronomisches Recheninstitut, Philosophenweg 12, 
69120 Heidelberg, Germany\\
$^{2}$ Jodrell Bank Centre for Astrophysics, School of Physics and Astronomy, The University of Manchester, Manchester, 
M13 9PL, United Kingdom\\
$^{3}$Zentrum f{\"u}r Astronomie der Universit{\"a}t Heidelberg, Institut f{\"u}r theoretische Astrophysik, 
Philosophenweg 12, D-69120, Heidelberg, Germany}

\date{Accepted ?, Received ?; in original form \today}

\pagerange{\pageref{firstpage}--\pageref{lastpage}} \pubyear{2016}

\begin{document}

\label{firstpage}

\maketitle

% -------------------------------------------------------------------- %
\begin{abstract}
We study the spherical collapse model in the presence of external gravitational tidal shear fields for different dark 
energy scenarios and investigate the impact on the mass function and cluster number counts. While previous studies of 
the influence of shear and rotation on $\delta_\mathrm{c}$ have been performed with heuristically motivated models, we 
try to avoid this model dependence and sample the external tidal shear values directly from the statistics of the underlying linearly evolved density field based on first order Lagrangian perturbation theory. 
Within this self-consistent approach, in the sense that we restrict our treatment to scales where linear theory is 
still applicable, only fluctuations larger than the scale of the considered objects are included into the sampling 
process which naturally introduces a mass dependence of $\delta_\text{c}$. We find that shear effects are predominant 
for smaller objects and at lower redshifts, i.~e. the effect on $\delta_\mathrm{c}$ is at or below the percent level 
for the $\Lambda$CDM model. 
For dark energy models we also find small but noticeable differences, similar to $\Lambda$CDM. The virial overdensity 
$\Delta_\mathrm{V}$ is nearly unaffected by the external shear. The now mass dependent $\delta_\mathrm{c}$ is used to 
evaluate the mass function for different dark energy scenarios and afterwards to predict cluster number counts, 
which indicate that ignoring the shear contribution can lead to biases of the order of $1\sigma$ in the estimation of 
cosmological parameters like $\Omega_\text{m}$, $\sigma_8$ or $w$.
\end{abstract}


% --------------------------------------------------------------------%
%---------------------------------------------------------------------%

\section{Introduction}\label{sec:1}
Since a decade cosmological observations provide very tight constraints on the parameters allowed within a certain 
class of models. With this ever increasing precision it is necessary to provide robust and accurate model predictions 
for future experiments. Combined observations of type-Ia supernovae \citep[e.g.][]{Riess1998,Perlmutter1999}, the
cosmic microwave background \citep[e.g.][]{Komatsu2011,Planck2015_XIII}, the Hubble constant and large-scale structure 
\citep[e.g.][]{Cole2005} show that the universe is spatially flat and expanding in an accelerated fashion. Assuming the 
symmetries of standard cosmology and General Relativity to be true, the accelerated expansion can be described by the 
cosmological constant $\Lambda$ or by introducing a fluid component, dubbed dark energy 
\citep[see e.g.][for a review]{Copeland2006}, with an equation of state $w<-1/3$, which in principle can vary with 
time. The cosmological constant corresponds to a constant equation of state $w=-1$. So far there is no significant 
evidence for any departure from $\Lambda$.

However, even though the constraints on $w$ are quite tight today, its time evolution is constrained rather poorly. 
Therefore it is possible to allow for temporal variations in the equation of state. Models described by a generic 
time-varying equation of state are referred to as dynamical dark energy models. Despite the intense theoretical effort 
in revealing the nature of dark energy, the fundamental origin of dark energy is still unknown, therefore the majority 
of the studies are based on phenomenological assumptions on the time evolution of the dark energy equation of state. 
Once this quantity is specified, all the properties of dark energy at the background level are known. From a 
theoretical point of view, time-varying equations of state can naturally be achieved within the framework of scalar 
fields. In these models, once their self-interaction potential is specified, the time evolution of the scalar field is 
obtained by solving a Klein-Gordon equation and as consequence also the corresponding equation of state can be 
evaluated. Under the generic term of scalar field models, we have many sub-classes, such as quintessence models, 
phantom models, $k$-essence, tachyon models and so forth. Quintessence and phantom models can be accommodated within 
the minimally-coupled model class and the equation of state can be either strictly greater than -1 (quintessence) or 
smaller (phantom). Dark energy models do not affect only the background evolution by changing the Hubble factor, but 
also the evolution of structures. 
In addition, even if sub-dominant, dark energy can possess perturbations for $w\ne -1$.

A promising tool to reveal the time evolution of dark energy observationally is the halo mass function, which enters for 
example in cluster counts \citep{Sunyaev1980b,Majumdar2004,Diego2004,Fang2007,Abramo2009a,Angrick2009} or weak lensing 
peak counts \citep{Maturi2010,Maturi2011,Lin2014,Reischke2016}. The halo mass function deals with objects in the 
highly non-linear regime and therefore a method is needed to extrapolate the linearly evolved density to the 
non-linear one. 
This is usually done by using the spherical collapse model introduced by \citet{Gunn1972} and later extended in several works 
\citep{Fillmore1984,Bertschinger1985,Ryden1987,AvilaReese1998,Mota2004,Abramo2007,Pace2010,Pace2014}. 
The model assumes perturbations to be spherically symmetric non-rotating objects which decouple from the background 
expansion and thus reach a maximum point of expansion after which they collapse. In principle they would collapse to a 
single point. However, in reality  the kinetic energy due to the collapse is converted into random 
motions of the particles in the over-dense regions, such that an equilibrium situation 
\citep[in the sense of virialized structure,][]{Schafer2008} is created. This model is, despite its simplicity, rather 
successful.

It is therefore important to get some insight into the theoretical assumptions of this model and to extend it towards 
more realistic situations. Especially rotation and shear effects are important extensions to the collapse model. Mainly 
rotational effects have been described in \citet{Pace2014b} which delay the collapse due to centrifugal forces, thus 
delaying the collapse of structures leading to a larger over-density needed for virialized structures. As the collapse 
model assumes a homogeneous sphere, shear effects are usually neglected, however, there can also be shear effects in 
homogeneous spheres and as real structures form in over-dense regions, there there will be shear effects due to external tidal 
fields. Those, if small enough, would not violate the symmetry assumptions of the model. External shear automatically 
leads to a mass dependence of the fundamental parameter of the spherical collapse, the critical over-density 
$\delta_\text{c}$, as light and therefore smaller objects will feel higher fluctuations in the density field than heavy 
objects. 
In this paper we will investigate the influence of external shear effects and how it depends on the underlying cosmological 
model. To this end we calculate the shear directly from the underlying density field by using first order Lagrangian 
perturbation theory, i.e. the Zel'dovich approximation 
\citep{zeldovich1970}. We set up a random process to sample shear values from the statistics of the underlying density 
field and investigate how this affects the collapse on different scales. This procedure has the advantage that we do 
not need to rely on phenomenological models, as we can instead calculate the tidal shear from first principles as it is 
for example also done in angular momentum correlations of large scale structure due to tidal torquing 
\citep{Schafer2009}.

The structure of the paper is as follows. In \autoref{sec:SC} we review the spherical collapse model and show the 
equations to be solved. In \autoref{sec:Tidal} we introduce a statistical procedure to obtain tidal shear values for 
the collapse. This method is then used in \autoref{sec:dcLCDM} to calculate the influence of the tidal shear on 
$\delta_\text{c}$ for the standard $\Lambda$CDM model which is later generalized to more complicated dark energy models 
in \autoref{sec:DECDM}. In \autoref{sec:MF} and \ref{sec:clustercounts} we investigate the influence of shear effects 
on the mass function and on cluster counts due to the Sunyaev-Zel'dovich effect and how a negligence of tidal shear 
effects can bias measurements of cosmological parameters. We summarize our findings in \autoref{sec:concl}.

% -------------------------------------------------------------------- %
\begin{figure}
 \begin{center}
  \includegraphics[width=0.45\textwidth]{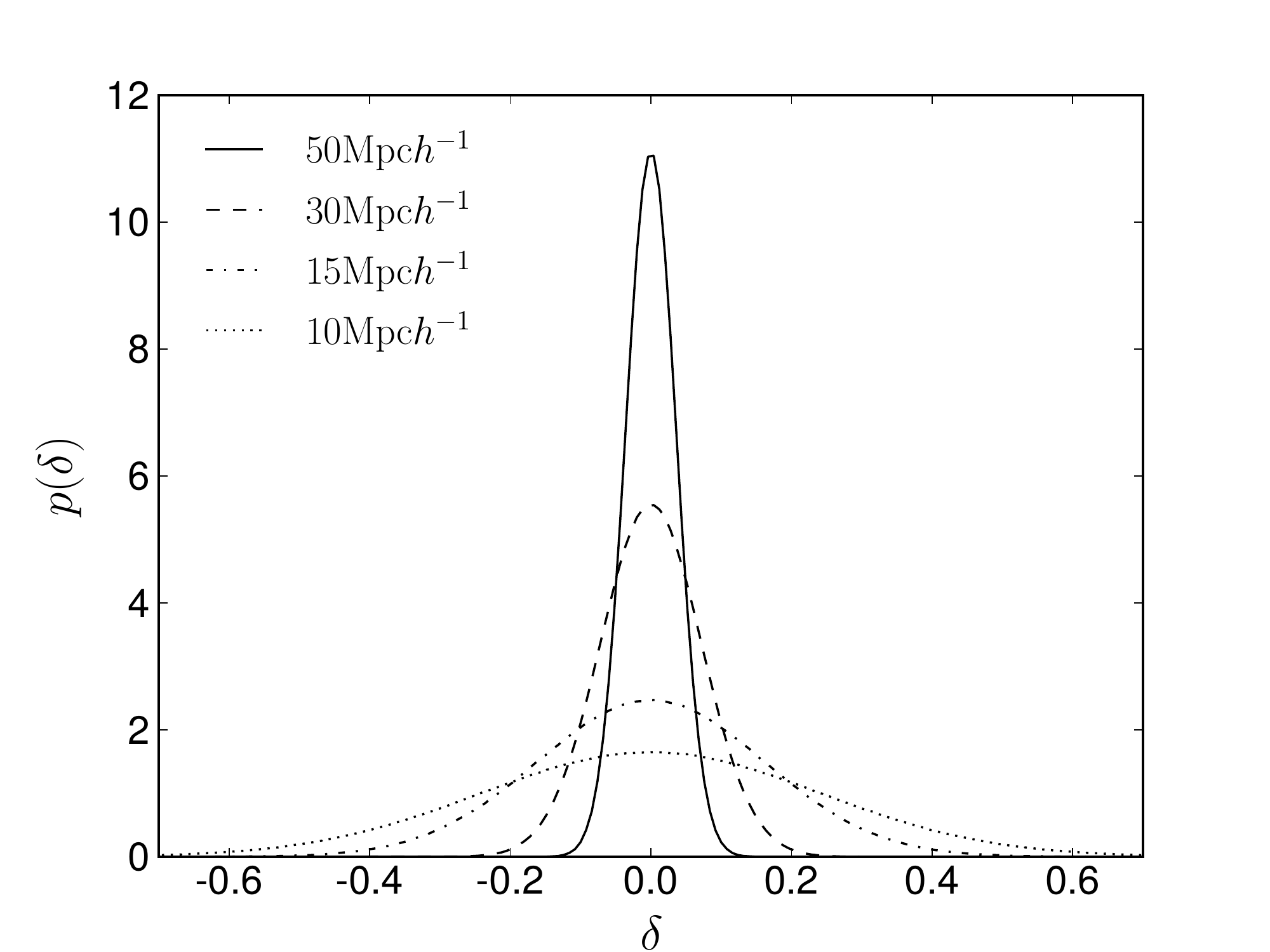}
  \caption{Normalized distribution of the density contrast $\delta = \psi_{ii}$ for different length scales. 
  Note that we show the distribution of $\sigma_\text{s}^2$ with an offset of unity on the right. 
  Clearly the values for $\delta$ below $R\approx 10\, \text{Mpc}h^{-1}$ would become too large in order to satisfy 
  the assumption $\delta\ll 1$.}
  \label{Fig:1}
 \end{center}
\end{figure}

\begin{figure}
 \begin{center}
  \includegraphics[width=0.44\textwidth]{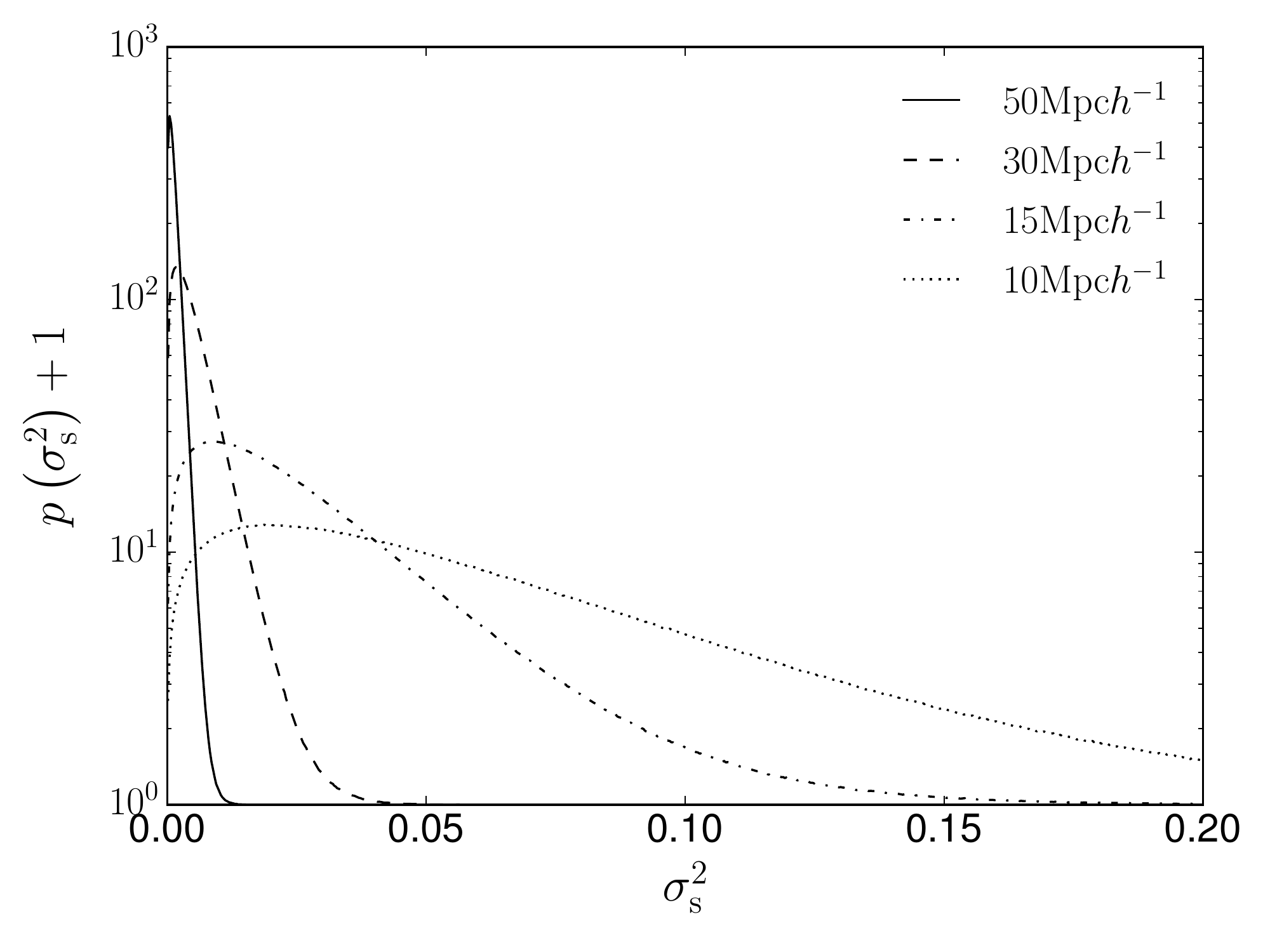}
  \caption{Normalized distribution of the tidal shear invariant 
  $\sigma_\text{s}^2$ given in Eq.~(\ref{eq:shear}) for different length scales. 
  Note that we show the distribution of $\sigma_\text{s}^2$ with an offset of unity on the right.}
  \label{Fig:sigmas}
 \end{center}
\end{figure}

\begin{figure*}
 \begin{center}
  \includegraphics[width = 0.45\textwidth]{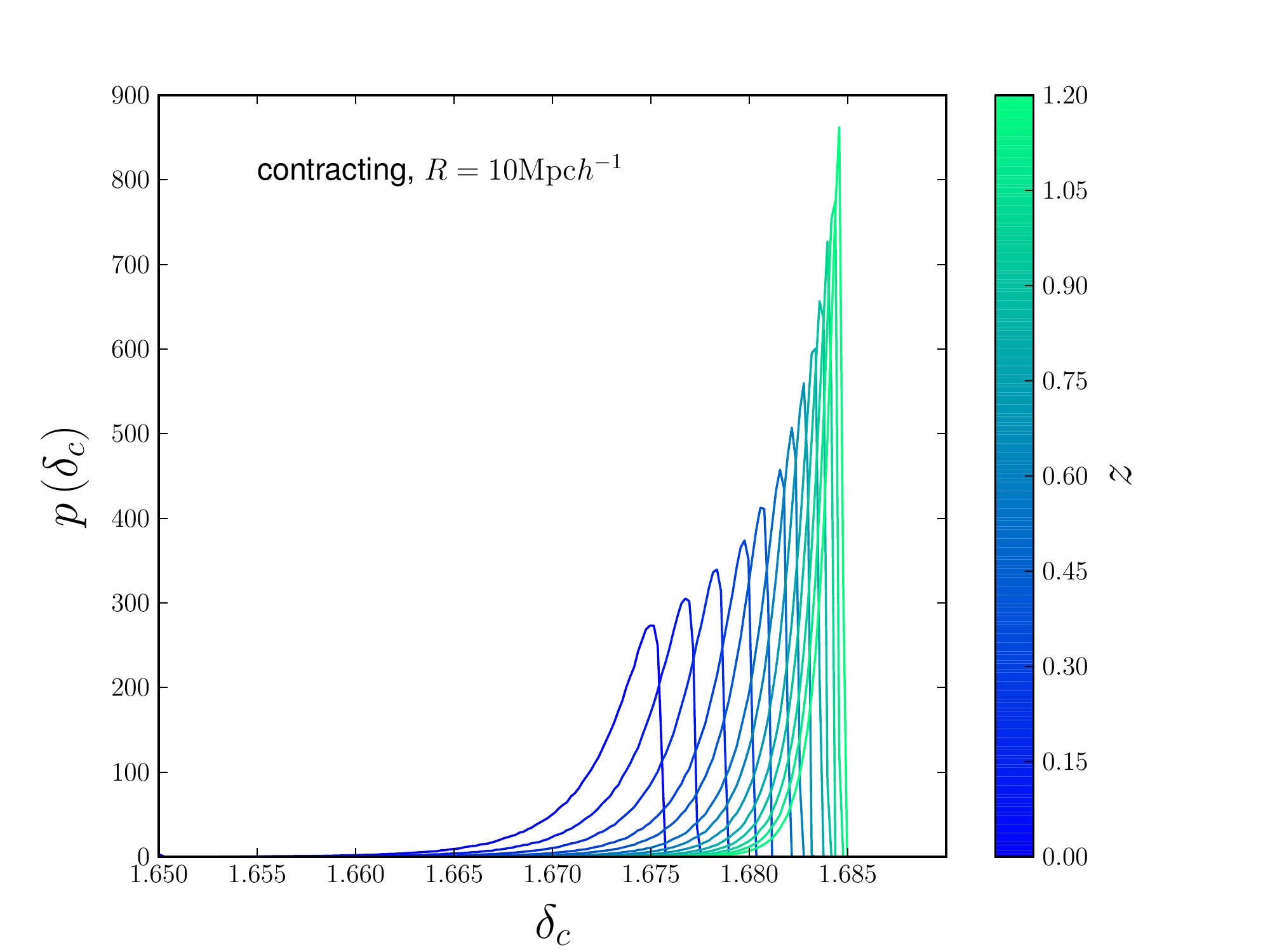}
  \includegraphics[width = 0.45\textwidth]{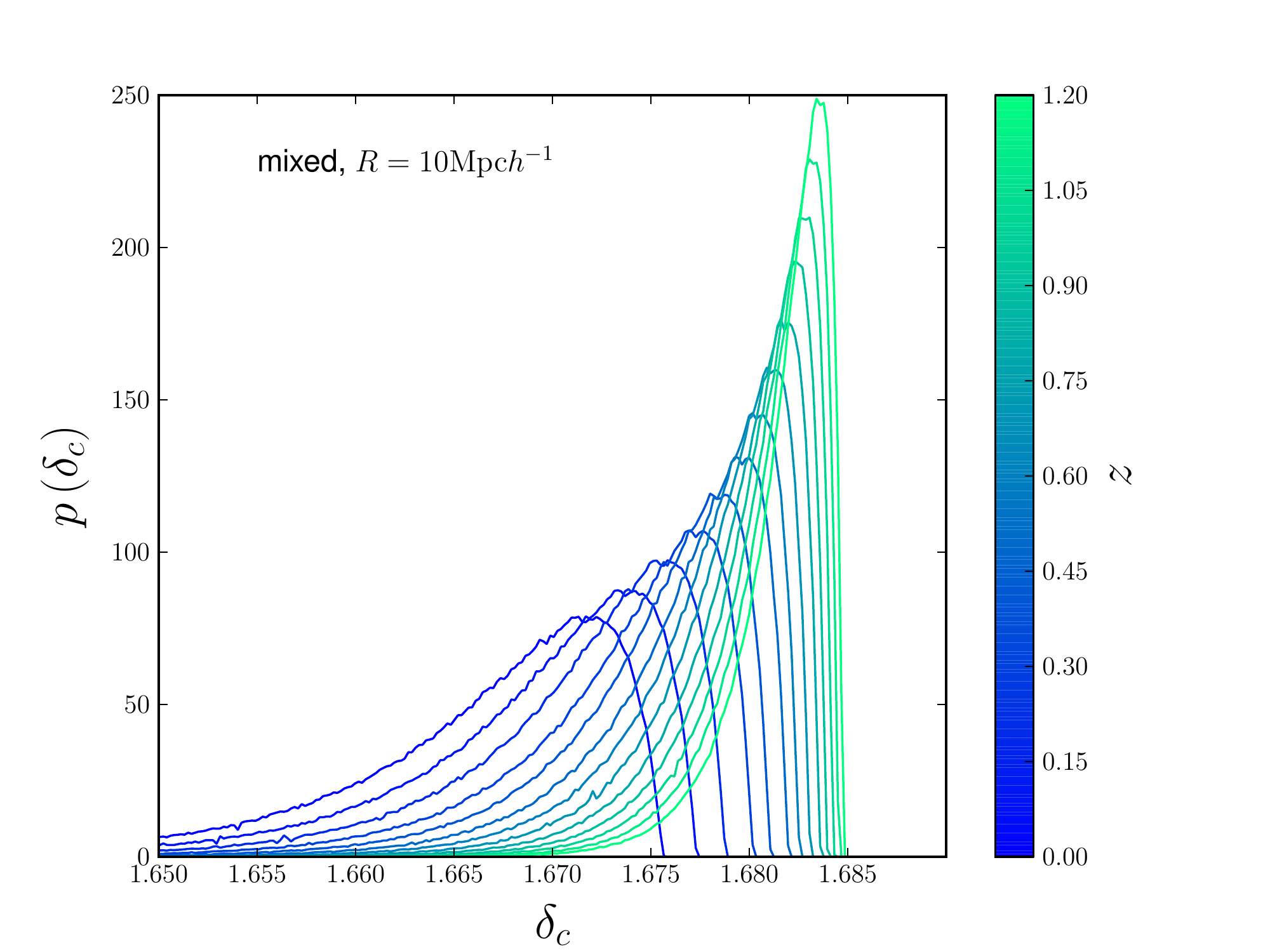}
  \caption{Distribution of $\delta_{\rm c}$ in over-dense regions at different redshifts for a radius 
  $R= 10\, \text{Mpc}h^{-1}$. \textit{Left}: Contracting regions, i.e. regions with only positive eigenvalues of 
  the velocity divergence tensor. \textit{Right}: Regions which are partially expanding and partially 
  contracting.}
  \label{Fig:5}
 \end{center}
\end{figure*}

% -------------------------------------------------------------------- %
% -------------------------------------------------------------------- %
\section{The Spherical Collapse Model}\label{sec:SC}
The spherical collapse model has been discussed by various authors, e.g 
\citet{Bernardeau1994,Padmanabhan1996,Ohta2003,Ohta2004,Abramo2007} and \citet{Pace2010,Pace2014}. 
Here we start with the hydrodynamical equations
\begin{equation}\label{eq:1}
 \begin{split}
  \dot{\delta} + (1+\delta)\nabla_{\boldsymbol x}\boldsymbol u = & \ 0\;, \\ 
  \dot{\boldsymbol u} + 2H\boldsymbol u+ (\boldsymbol u\cdot\nabla_{\boldsymbol x})\boldsymbol u = & \ 
  -\frac{1}{a^2}\nabla_{\boldsymbol x}\phi \;, \\
  \nabla^2_{\boldsymbol x}\phi = & \ 4\pi G a^2\rho_0 \delta\;,
 \end{split}
\end{equation}
with comoving coordinate $\boldsymbol x$, comoving peculiar velocity $\boldsymbol u$, Newtonian potential $\phi$, 
overdensity $\delta$ and background density $\rho_0$. The dot represents a derivative with respect to cosmic time $t$. 
Taking the divergence of the Euler equation and inserting the Poisson equation yields 
\begin{equation}\label{eq:2}
 \begin{split}
  \dot\delta = & -(1+\delta)\theta\;, \\
  \dot\theta = & -2H\theta -4\pi G\rho_0 \delta-\frac{1}{3}\theta^2 -(\sigma^2-\omega^2)\;,
 \end{split}
\end{equation}
where we used the decomposition
\begin{equation}\label{eq:3}
 \nabla_{\boldsymbol x} \cdot [(\boldsymbol u\nabla_{\boldsymbol x}) \boldsymbol u)] = 
 \frac{1}{3}\theta^2 +\sigma^2-\omega^2\;,
\end{equation}
with the expansion $\theta = \nabla_{\boldsymbol x}\cdot\boldsymbol u$, the shear 
$\sigma^2 \equiv \sigma_{ij}\sigma^{ij}$ and the rotation $\omega^2\equiv \omega_{ij}\omega^{ij}$. The rotation and the 
shear tensors are themselves the antisymmetric and the symmetric traceless part of the velocity divergence tensor, 
respectively. They are defined as
\begin{equation}
\begin{split}
\sigma_{ij} = & \ \frac{1}{2}\left(\partial_i u_j +\partial_j u_i\right) - \frac{\theta}{3}\delta_{ij} \\
\omega_{ij} = &  \ {\frac{1}{2}}\left(\partial_i u_j -\partial_j u_i\right),
\end{split}
\end{equation}
where $\partial_i\equiv \partial/\partial x^i$. 
We now use the relation $\partial_t = aH(a)\partial_a$ and $f\equiv 1/\delta$ which leads to
\begin{equation}\label{eq:4}
 \begin{split}
 f^{\prime} = & \ \frac{\theta}{aH}f(1+f)\;, \\
 \theta^{\prime} = & -\frac{2\theta}{a}-\frac{3H\Omega_{\text{m}}}{2af}-
 \left(\frac{1}{3}\theta^2+\sigma^2-\omega^2\right)\frac{1}{aH}\;.
 \end{split}
\end{equation}
The system in Eq.~(\ref{eq:4}) is solved numerically until $f\sim 10^{-14}$ and then it is extrapolated to zero. This 
yields the appropriate initial conditions for the linear evolution of the density contrast which gives 
$\delta_{\rm c}$. 
In the classical spherical collapse model, $\sigma^2$ and $\omega^2$ are neglected. However, their influence has been 
investigated by \citet{DelPopolo2013a,DelPopolo2013b} in the $\Lambda$CDM and dark energy cosmologies and by 
\cite{Pace2014b} in clustering dark energy models. The authors employ a heuristic model for the term 
$\sigma^2-\omega^2$ which allows to study an isolated collapse including a (mass dependent) quantity $\alpha$, defined 
as the ratio between the rotational and the gravitational term. Quantitatively, the term is
\begin{equation}
 \alpha=\frac{L^2}{M^3RG}\;,
\end{equation}
where $L$ denotes the angular momentum of the spherical overdensity considered and $M$ and $R$ its mass and radius, 
respectively. The angular term is important for galaxies and negligible for massive clusters; in particular 
$\alpha\approx 0.05$ for $M\approx 10^{11}~{\rm M}_{\odot}~h^{-1}$ and of the order of $10^{-6}$ for 
$M\approx 10^{15}~{\rm M}_{\odot}~h^{-1}$. By defining the twiddled quantities $\tilde{\theta}=\theta/H$, 
$\tilde{\sigma}=\sigma/H$ and $\tilde{\omega}=\omega/H$, the combined contribution of the shear and rotation term can 
effectively be modeled by
\begin{equation}
 \tilde{\sigma}^2-\tilde{\omega}^2=-\frac{3}{2}\alpha\Omega_{\rm m}\delta\;,
\end{equation}
leading to the modified Euler equation
\begin{equation}\label{eq:euler_mod}
 \tilde{\theta}^{\prime}+\left(\frac{2}{a}+\frac{H^{\prime}}{H}\right)\tilde{\theta}+\frac{\tilde{\theta}^2}{3a}+
 \frac{3}{2a}(1-\alpha)\Omega_{\rm m}\delta=0\;.
\end{equation}
In the notation of this work, Eq.~(\ref{eq:euler_mod}) reads
\begin{equation}
 \tilde{\theta}^{\prime}+\left(\frac{2}{a}+\frac{H^{\prime}}{H}\right)\tilde{\theta}+\frac{\tilde{\theta}^2}{3a}+
 \frac{3}{2a}\frac{(1-\alpha)\Omega_{\rm m}}{f}=0\;.
\end{equation}
As shown by the authors, the effect of the term $\tilde{\sigma}^2-\tilde{\omega}^2$ is to slow down the collapse and to 
decrease the number of objects. This effect is differential and depends on mass and on redshift. At high redshifts, 
modifications are small, while at low redshifts they are more substantial. 
In addition, we can appreciate the slowing of the collapse (now mass dependent) for low mass objects.

In this work we follow a complementary approach. Instead of trying to model the additional non-linear term, we will 
derive only the shear contribution from the statistics of the density field in linear perturbation theory, since at 
early times velocities decay rapidly and vorticity is not sourced in the linear regime. Hence a direct comparison with 
the work by \cite{DelPopolo2013a,DelPopolo2013b,Pace2014b} cannot be performed. Note, however that we can expect an 
opposite behaviour of the collapse, since it is well known \citep{Angrick2010} that the ellipsoidal collapse proceeds 
faster than the spherical collapse.

\begin{figure}
 \begin{center}
  \includegraphics[width = 0.45\textwidth]{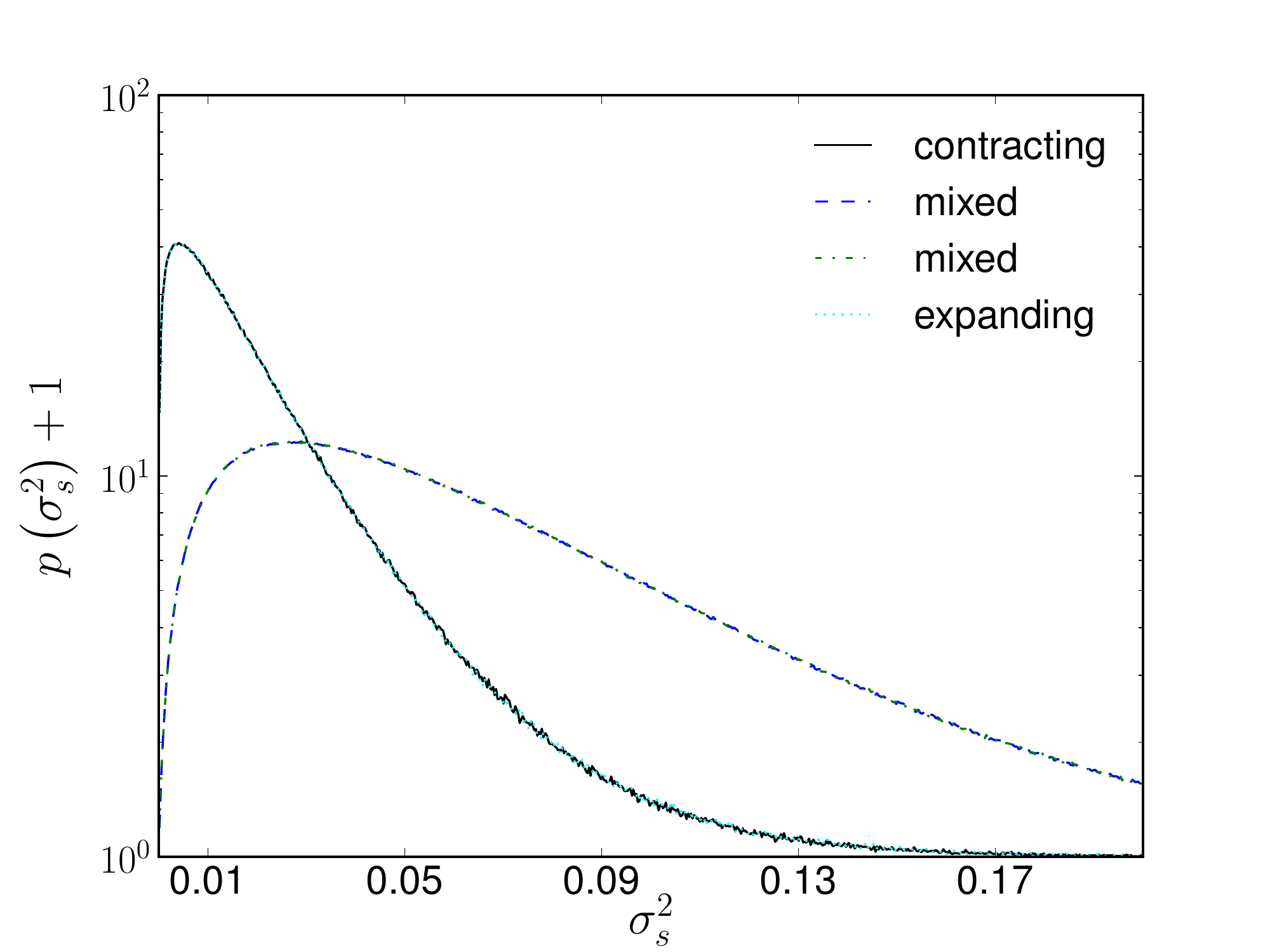}
  \caption{Distribution of the shear invariant $\sigma_\text{s}^2$ for different environments. We found 
  that the mixed environment, i.e. where $\sigma_{ij}$ has positive and negative eigenvalues, is much more likely than 
  the contracting or expanding environment making up for approximately $95\%$ of the sampled values. 
  Note that we again show the normalized distribution with an offset of unity.}
  \label{Fig:3}
 \end{center}
\end{figure}

\begin{figure*}
 \begin{center}
  \includegraphics[width = 0.45\textwidth]{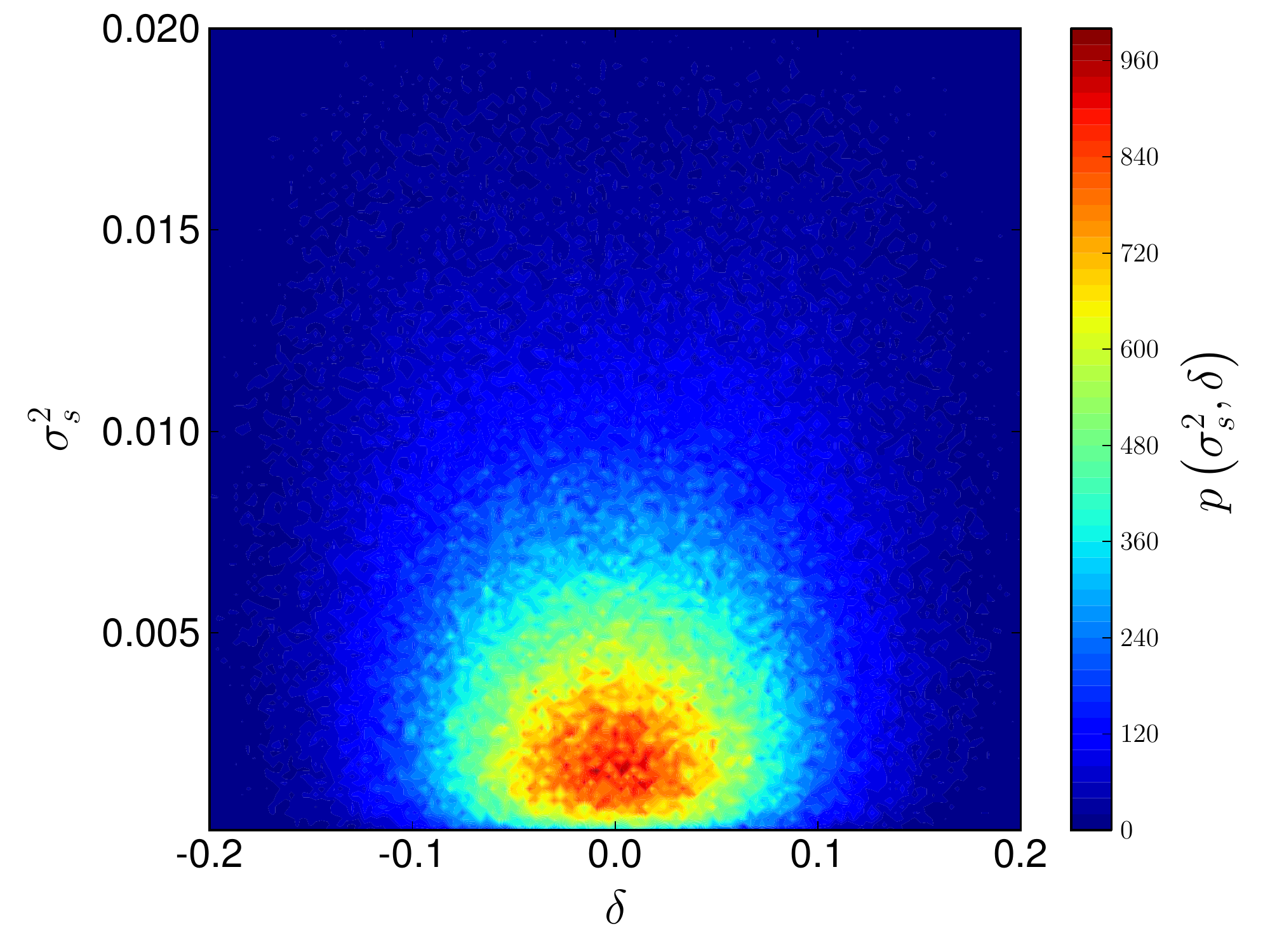}  
  \includegraphics[width = 0.45\textwidth]{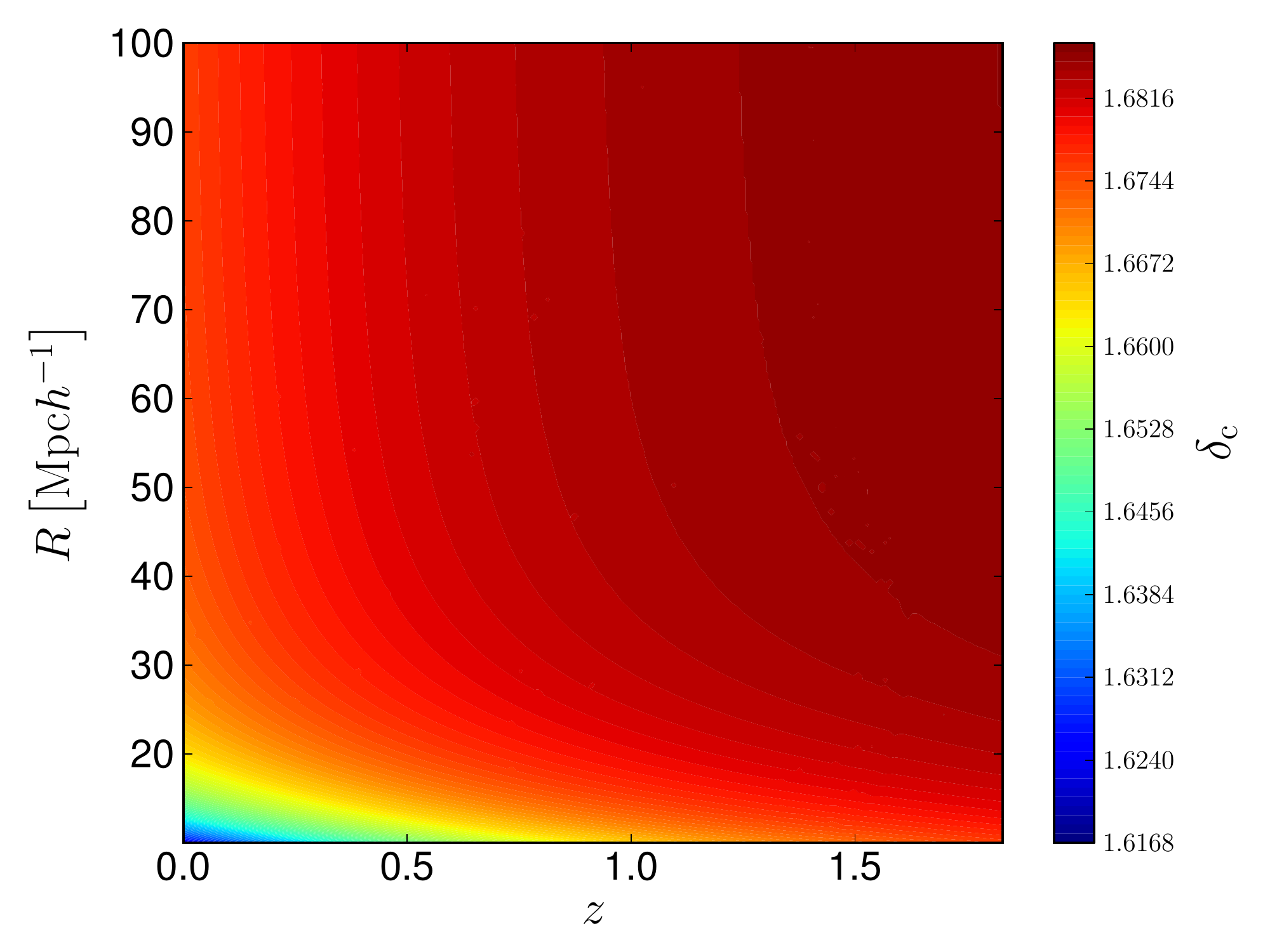}
  \caption{\textit{Left panel}: Joint distribution of $\delta$ and $\sigma_\text{s}^2$ for a smoothing radius 
  $R=30\;\text{Mpc}h^{-1}$. 
  \textit{Right panel}: Mean linearly evolved critical over-density $\bar{\delta}_\text{c}(M)$ including the shear 
  as a function of redshift and object scale which is related to the mass of an object via 
  $M =\frac{4\pi}{3}R^3\rho_0$, with 
  $\rho_0 =\rho_\text{crit}\Omega_\text{m}$. Clearly the effect of external shear is most pronounced at low 
  redshifts and low masses, while it converges to the standard $\Lambda$CDM value for the other cases.}
  \label{Fig:6}
 \end{center}
\end{figure*}

\section{Sampling Tidal Shear Values}\label{sec:Tidal}
For the tidal shear we assume Zel'dovich velocities \citep{zeldovich1970}, thus approximating the velocity field as a 
potential flow. For the trajectories one assumes
\begin{equation}\label{eq:Zeld}
 x_i = q_i-D_+(t)\partial_i \psi \equiv q_i -D_+(t)\psi_{,i}\;,
\end{equation}
with the displacement field $\psi$ which is related to the density contrast $\delta$ via a Poisson relation, 
$\Delta \psi = \delta$, the initial position $q$ and the linear growth factor $D_+(t)$. The velocity is then given by
\begin{equation}\label{eq:6}
 \dot x_i(t) = -\dot{D}_+(t)\psi_{,i} = -H\frac{\text{d}\ln D_+}{\text{d}\ln a}D_+\psi_{,i}\;.
\end{equation}
Clearly there is no vorticity in this configuration, due to the permutability of the second derivatives. Thus the only 
remaining contribution to the spherical collapse is the traceless shear tensor
\begin{equation}\label{eq:shear}
 \sigma^2 \equiv \sigma_{ij}\sigma^{ij} =\dot{D}_+^2(t)\left(\psi_{,ij}\psi^{,ij}-\frac{1}{3}(\Delta\psi)^2\right) 
 \equiv \dot{D}_+^2(t)\sigma_\text{s}^2\;,
\end{equation}
with $\psi_{,ij}\equiv\partial_i\partial_j\psi$. In the last step the time evolution was separated from the constant 
shear $\sigma_\text{s}^2$. We now sample values for the shear, $\psi_{,ij}$ directly from the statistics of the 
underlying density field. To this end we transform to Fourier space and use Poisson's equation leading to
\begin{equation}
 \psi_{,ij} = \int\frac{\text{d}^3k}{(2\pi)^3}\frac{k_ik_j}{k^2}\delta(\boldsymbol k)\exp(\text{i}\boldsymbol{k}
              \boldsymbol{x})\;.
\end{equation}
However, the correlation between the density field and the tidal shear is complicated in these coordinates. Following 
\citet{Regos1995} and \citet{Heavens1999} we consider the density peaks symmetric about the origin on the $z$-axis and 
introduce dimensionless complex variables
\begin{equation}
 y^n_{lm} = \sqrt{4\pi}\frac{\text{i}^{l+2n}}{\sigma_{l+2n}}\int\frac{\text{d}^3k}{(2\pi)^3}k^{l+2n}
 \delta(\boldsymbol k) Y_{lm}(\hat{k})\exp(\text{i}\boldsymbol k\boldsymbol x)\;,
\end{equation}
with the direction vector $\hat{k}=\boldsymbol k/k$ and $\sigma_i$ being the spectral moments of the matter power 
spectrum
\begin{equation}
 \sigma_i^2 = \frac{1}{2\pi^2}\int\text{d}k\; k^{2i+2} P(k),
\end{equation}
while $Y_{lm}$ are spherical harmonics. We obtain a linear relation \citep{Schafer2012} between $y^n_{lm}$ 
and the tidal shear values $\psi_{,ij}$
\begin{equation}\label{eq:11}
\begin{split}
\sigma_0 y_{20}^{-1} = & \ -\sqrt{\frac{5}{4}}\left(\psi_{,xx}+\psi_{,yy}-2\psi_{,zz}\right)\;, \\
\sigma_0 y^{-1}_{2\pm 1} = & \ -\sqrt{\frac{15}{2}}\left(\psi_{,xz}\pm\text{i}\psi_{,yz}\right)\;, \\
\sigma_0 y^{-1}_{2\pm 2} = & \ \sqrt{\frac{15}{8}}\left(\psi_{,xx}-\psi_{,yy}\pm2\text{i}\psi_{,xy}\right)\;, \\
\sigma_0 y^{0}_{00} = & \ \left(\psi_{,xx}+ \psi_{,yy}+\psi_{,zz}\right)\;.
\end{split} 
\end{equation}
In particular, the covariance in this basis is trivial, since the auto-correlation matrix is diagonal in $l$ and $m$:
\begin{equation}
 \left\langle y_{lm}^n(\boldsymbol x)y_{l'm'}^{n'}(\boldsymbol x)^*\right\rangle =(-1)^{n-n'}\frac{\sigma^2_{l+n+n'}}
 {\sigma_{l+2n}\sigma_{l+2n'}}\delta_{ll'}\delta_{mm'}\;.
\end{equation}
Thus, in the $y^n_{lm}$ basis the tidal shear values are uncorrelated Gaussian random variables with unit variance. We 
obtain the tidal shear values in physical coordinates by inverting the mapping
\begin{equation}
 \sigma_0\boldsymbol\alpha = M\boldsymbol \psi\;,
\end{equation}
where the six dimensional vectors $\boldsymbol\alpha$ and $\boldsymbol\psi$ bundle the variables in spherical and 
physical coordinates from Eq. (\ref{eq:11}) respectively
\begin{equation}\begin{split}
 \boldsymbol\alpha^\text{T} = & \ \left(y_{00}^0,y_{20}^{-1},y_{21}^{-1},y_{2-1}^{-1},y_{22}^{-1},y_{2-2}^{-1}
                                  \right)\;,\\
 \boldsymbol\psi^\text{T} = & \ \left(\psi_{,xx},\psi_{,yy},\psi_{,zz},\psi_{,xy},\psi_{,xz},\psi_{,yz}\right)\;.
\end{split}\end{equation}
The inverse mapping $M^{-1}$ is then given by
\begin{equation}
 M^{-1}= \begin{pmatrix}
 1/3 & -\frac{\sqrt{5}}{15} & 0 & 0 & \frac{\sqrt{30}}{30} & \frac{\sqrt{30}}{30} \\
 1/3 & -\frac{\sqrt{5}}{15} & 0 & 0 & -\frac{\sqrt{30}}{30} & -\frac{\sqrt{30}}{30}\\ 
 1/3 & 2\frac{\sqrt{5}}{15} & 0 & 0 & 0 & 0 \\
 0 & 0 & 0 & 0 &-\frac{\sqrt{30}}{30}\text{i} & \frac{\sqrt{30}}{30}\text{i}\\
 0 & 0  &-\frac{\sqrt{30}}{30}  &-\frac{\sqrt{30}}{30} & 0 & 0 \\
 0 & 0  &\frac{\sqrt{30}}{30}\text{i}  &-\frac{\sqrt{30}}{30}\text{i} & 0 & 0
 \end{pmatrix}
\end{equation}
Note that the components $y^i_{l\pm m}$ are Hermitian conjugate variables, thus preserving the real nature of the shear 
field. The amount of tidal shear acting on a halo depends on the length scale $R(M)$ of the halo and thus on its mass. 
In our model a halo will only be affected by the shear caused by structures with length scale $L\gtrsim R(M)$. 
Therefore we introduce a cut-off for the power spectrum, suppressing high frequencies
\begin{equation}
 P(k) \to P(k)W^2_R(k), 
\end{equation}
with $W_R(k) = \exp(-k^2R^2/2)$. The mass scale is obtained via 
$M = \frac{4\pi}{3}\rho_\text{crit}\Omega_\text{m} R^3$, where $\rho_\text{crit}= 3H^2/(8\pi G)$ is the critical 
density. Here all quantities are evaluated today, as the time dependence is taken into account via the time derivative 
of the growth factor in Eq.~(\ref{eq:Zeld}). From the sampled shear values $\psi_{,ij}$ the shear invariant $\sigma^2$ 
can be calculated using Eq.~(\ref{eq:shear}).

Clearly for low mass haloes shear becomes more important as the fluctuations in the surrounding density field are 
larger. Since our model works with a potential flow for the velocities, the variance $\sigma_0$ must remain small 
compared to $|\delta|=1$, showing the validity of the treatment presented here above a certain scale only on which the 
evolution of the 
density contrast can safely be considered as linear. In \autoref{Fig:1} we show the distribution of the sampled density 
contrast $\delta = \psi_{ii}$ for different mass scales. It is 
easy to see that smoothing of the density 
field on smaller scales leads to a broader distribution of delta.
Especially this shows that $R\approx 10\,\text{Mpc}h^{-1}$ is the smallest scale at which the approximation used here 
is applicable as higher order terms will dominate the perturbative expansion. Consequently the velocity field will no 
longer be a potential flow. Conversely larger scales $R(M)$ will lower the values of $\sigma_\text{s}^2$, thus high 
mass halos will 
be less affected compared to low mass ones. The distribution of the remaining 
tidal shear invariant $\sigma_\text{s}^2$ (cf. Eq. (\ref{eq:shear}) for details), again for different scales, can be seen in \autoref{Fig:sigmas}.

In \autoref{Fig:3} we show how the invariant $\sigma^2_\text{s}$ distinguishes between different environments. These 
are 
classified by the characteristic of the shear tensor $\sigma_{ij}$. Due to $\Delta\psi = \delta$, positive eigenvalues 
correspond to a collapsing region, while negative eigenvalues correspond to an expanding region. The other two 
possibilities, i.e. one or two positive eigenvalues, correspond to a mix of both effects. Clearly the tidal shear 
invariant does not distinguish between contracting and expanding regions, as only the square of the traceless shear 
tensor enters into the collapse equation. The same is true for the mixed environments. 
Thus, a fully contracting environment has the same effect on the collapse as a fully expanding one. 
As halos form in over-dense regions, we are rather interested in a shear value provided the density contrast in this 
region satisfies $\delta>0$. It is important to note that no correlations enter into the model by conditionalizing the 
random process in such a way. 
The latter effect is shown in the right panel of \autoref{Fig:3} where the joint distribution of $\sigma_\text{s}^2$ 
can be seen to be symmetric around 
$\delta = 0$ for different values of $\sigma_\text{s}^2$ as expected from the Gaussian assumption and from 
\autoref{Fig:1}. 
It is therefore not harmful to neglect all values of the shear matrix which describe an under-dense region. 
Note that a halo can also form in a large under-dense region. Our results would, however, not be influenced by this effect as we work in the linear regime.

\section{Effect of mass and environment}\label{sec:dcLCDM}
The critical linear over-density $\delta_{\rm c}$ in a homogeneous sphere depends on the initial conditions for the 
linear equation. Those are derived from the fully non-linear equation which in principle includes shear and rotation 
effects. 
Within our model $\delta_\text{c}$ will be influenced by the surrounding shear which is encapsulated in the invariant 
$\sigma_\text{s}^2$. As we have seen in Sect.~\ref{sec:Tidal} the shear values are distributed randomly due to the 
underlying density field with amplitudes given by the considered scale. Consequently $\delta_\text{c}$ will also 
exhibit a distribution rather than a distinct value.

The distribution of $\delta_\text{c}$ for different collapse redshifts can be seen in~\autoref{Fig:5}. Contracting 
environments get less support by tidal shear than mixed environments which is due to the fact that the shear is larger 
if not all directions are contracting. The high end of both distributions falls off very rapidly which is due to the 
distribution $\sigma_\text{s}^2$ growing steeply towards $\sigma_\text{s}^2=0$. The zero point marks the value for $\delta_\text{c}$ obtained without tidal shear 
because $\sigma^2$ only enters as a positive contribution in Eq.~(\ref{eq:4}) and thus a non-vanishing shear will move 
the initial conditions for the linear equation to lower values of $\delta$ resulting in a smaller value for 
$\delta_\text{c}$. Furthermore the distribution of $\delta_\text{c}$ becomes narrower if the collapse redshift 
increases. This is due to the evolution of $\sigma^2$ with redshift: physically shear becomes more important with time 
due to the growth of the cosmic density field. Note that this effect occurs only for $z$ larger than $0.3$ since the 
time evolution of $\sigma^2$ in Eq.~(\ref{eq:6}) has a maximum at this redshift. It coincides with the time when the 
cosmological constant starts dominating the expansion of the universe, slowing down the growth of structures again.

Having evaluated the distribution of $\delta_\text{c}$ we can define an effective $\bar{\delta}_\text{c}$ which is 
taken to be the mean:
\begin{equation}
\bar{\delta}_\text{c}(M) = \int\text{d}\delta_\text{c}\; \delta_\text{c} \; p(\delta_\text{c},M).
\end{equation}
This mean value is now a function of the mass of the considered halo, which is carried by the amplitude of the density 
fluctuations on scales larger than the corresponding scale $R(M)$ of the halo and of the redshift via the collapse 
equation. \autoref{Fig:6} shows $\bar{\delta}_\text{c}$ as a function of the halo mass in units of $M_\odot h^{-1}$ 
and of the redshift. As expected from the previous discussion, $\delta_{\rm c}$ is mostly influenced at small radii and 
small redshifts as shear effects are most important in this regime.

\section{Effect of cosmology}\label{sec:DECDM}
Previous works on the effects of shear and rotation on the parameters of the spherical collapse model showed that the 
behaviour of these additional non-linear terms is mildly affected by the change of the background cosmological model. 
While overall their mutual combination had the same qualitative effect (increase in $\delta_{\rm c}$ and negligible 
effect at high masses), differences of the order of several percent appeared across different cosmological models 
considered. In this section we analyse the effects of dark energy on the linear extrapolated density parameter 
$\delta_{\rm c}$ and on the virial overdensity when we add the contribution of the shear field as outlined in the 
previous sections. The models here investigated have been explored before with the same purpose, albeit, as said 
before, a direct comparison is not possible at this stage. For more details on the models we refer the reader to 
\cite{Pace2010} for homogeneous dark energy and to \cite{Pace2014b} for clustering dark energy models. 
We will explore the effect of dark energy inhomogeneities in a following work.

\begin{figure}
 \centering
 \includegraphics[width=0.45\textwidth]{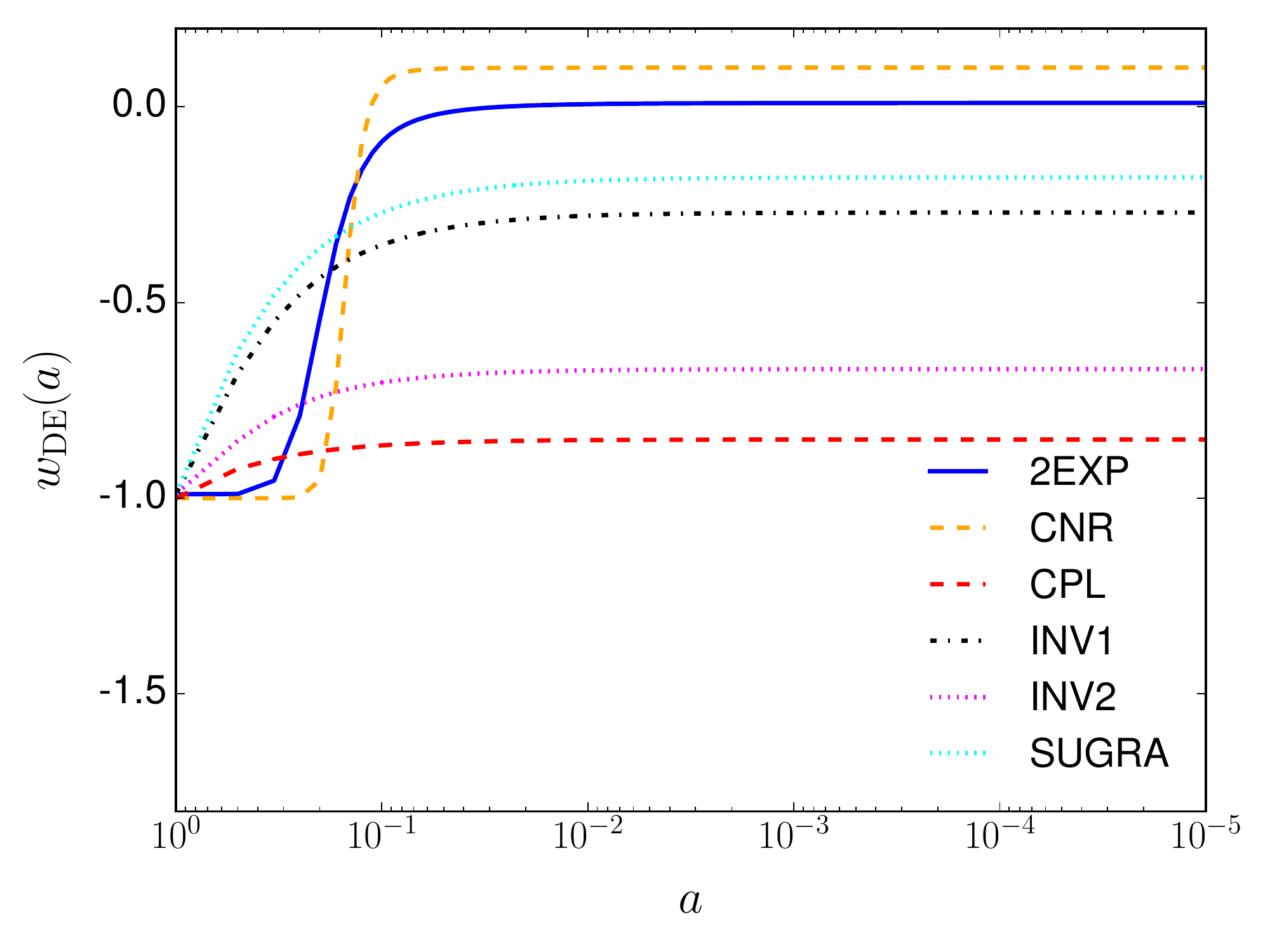}
 \caption{Time-dependent equations of state for the models used in this work as a function of the scale factor $a$. 
 The light-green dashed-dot black and the magenta short-dashed lines represent the model INV1 and INV2, respectively. 
 The blue curve the 2EXP model. The CPL and the CNR models are shown with the red dashed and the orange 
 dashed curve, respectively. Finally the SUGRA model is shown with the cyan dotted curve.}
 \label{fig:wa}
\end{figure}

In particular we will explore the effect of the tidal shear in models described by the following equation-of-state 
parametrization: three models with constant equation of state ($w_{\rm DE}=-1$ for the cosmological constant $\Lambda$, 
$w_{\rm DE}=-0.9$ for quintessence models and $w_{\rm DE}=-1.1$ for phantom models), and six models with a dynamical 
equation of state: 
\begin{itemize}
 \item the 2EXP model \citep{Barreiro2000},
 \item the CNR and the SUGRA model \citep{Copeland2000},
 \item the CPL model \citep{Chevallier2001,Linder2003},
 \item the INV1 and INV2 models \citep{Corasaniti2003,Corasaniti2004,Sanchez2009}.
\end{itemize}
In \autoref{fig:wa} we show for clarity the dynamical dark energy models used in this work. The CPL and the INV2 models 
show a very gentle increase of the equation-of-state parameter while the models SUGRA and INV1 present a more rapid 
change of the equation of state. 
The CNR model is approximately constant at low redshifts and is characterized by a sudden change for $a\approx 0.1$. 
All the models are approximately constant at small scale factors and $w_{\rm DE}\approx -1$ for $a\simeq 1$, as 
inferred from observational data.

The functional form for the CPL model is
\begin{equation}
 w_{\rm DE}(a)=w_0+w_\text{a}(1-a)\;,
\end{equation}
and we used $w_0=-1$ and $w_\text{a}=0.15$.

The other models can be well described by the following four-parameter formula:
\begin{equation}
 w_{\rm DE}=w_0+(w_{\mathrm{m}}-w_0)\frac{1+e^{\frac{a_{\mathrm{m}}}{\Delta_{\mathrm{m}}}}}
 {1+e^{-\frac{a-a_{\mathrm{m}}}{\Delta_{\mathrm{m}}}}}
 \frac{1-e^{-\frac{a-1}{\Delta_{\mathrm{m}}}}}{1-e^{\frac{1}{\Delta_{\mathrm{m}}}}}\;.
\end{equation}
In \autoref{tab:params} we summarize the values of the parameters used.

\begin{table}
 \caption{Parameter values for the dark energy models with dynamical equation-of-state parameter.}
 \begin{center}
  \begin{tabular}{c|c|c|c|c|c|}
   \hline
   \hline
   Model & $w_0$ & $w_{\mathrm{m}}$ & $a_{\mathrm{m}}$ & $\Delta_{\mathrm{m}}$ \\
   \hline
   2EXP  & -0.99 & 0.01 & 0.19 & 0.043 \\
   INV1  & -0.99 & -0.27 & 0.18 & 0.5 \\
   INV2  & -0.99 & -0.67 & 0.29 & 0.4 \\
   CNR   & -1.0 & 0.1 & 0.15 & 0.016 \\
   SUGRA & -0.99 & -0.18 & 0.1 & 0.7 \\
   \hline
  \end{tabular}
 \end{center}
 \label{tab:params}
\end{table}
Except for the EdS model where we assumed $\Omega_{\rm m}=1$, we will use for all the dark energy models the following 
set of parameters (assuming a flat spatial geometry): $\Omega_{\rm m}=0.32$, $\Omega_{\rm de}=0.68$, $h=0.67$ and 
$n_{\rm s}=0.966$.

\subsection{Spherical collapse parameters}\label{sec:spc}
In this section we will describe the effects of the introduction of the tidal shear on the two main parameters of the 
spherical collapse model: the linearly extrapolated overdensity $\delta_{\rm c}$ and the virial overdensity 
$\Delta_{\rm V}$. The first one is a very important theoretical quantity usually used in the determination of the mass 
function according to the prescription of \cite{Press1974} and \cite{Sheth1999}. 
The second one instead is used both in observations and in simulations to determine the mass and the size of the 
object. For details on how to evaluate them, we refer to \cite{Pace2010,Pace2012,Pace2014,Pace2014b}.

\begin{figure*}
 \centering
 \includegraphics[width=0.45\textwidth]{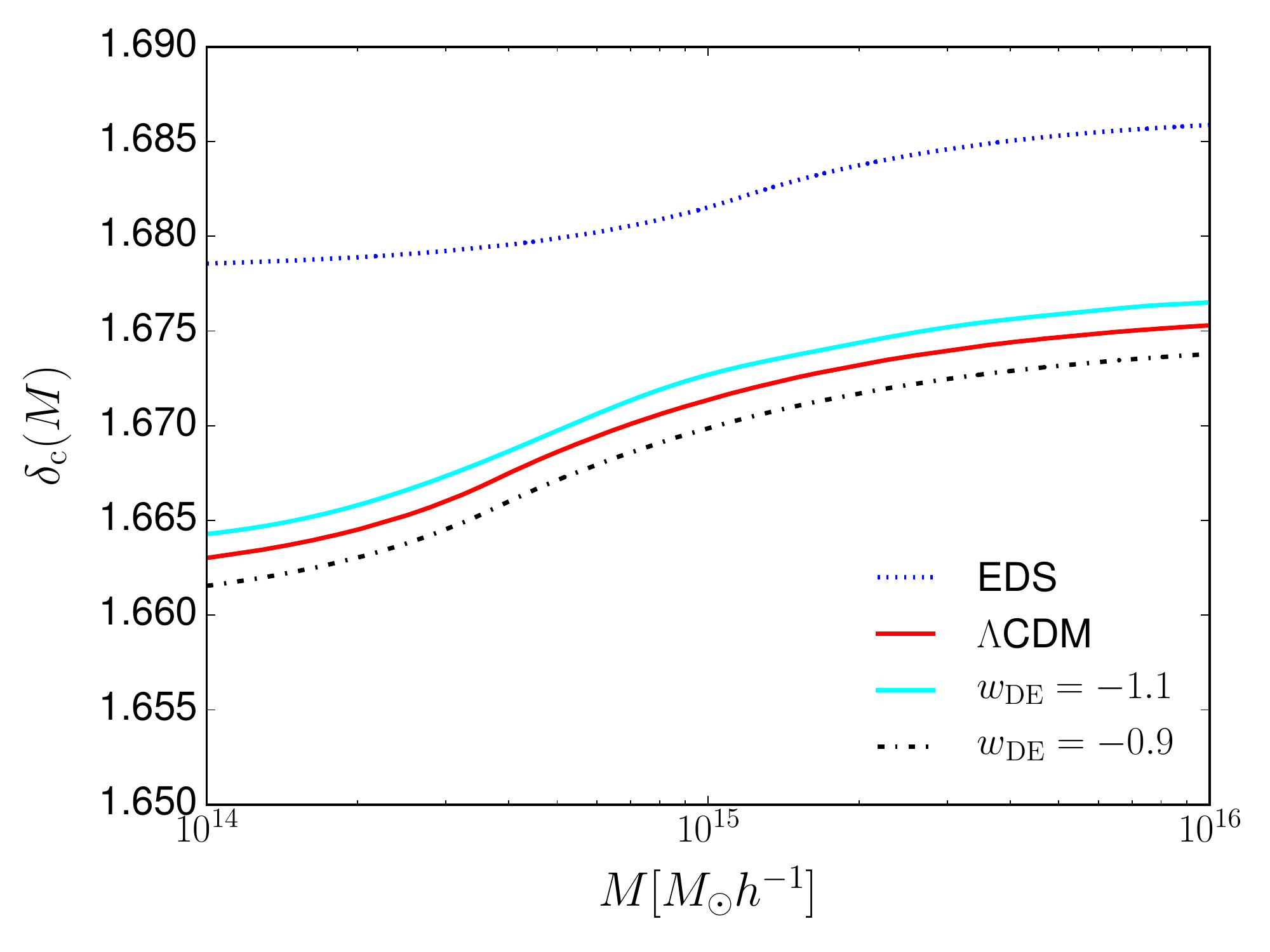}
 \includegraphics[width=0.45\textwidth]{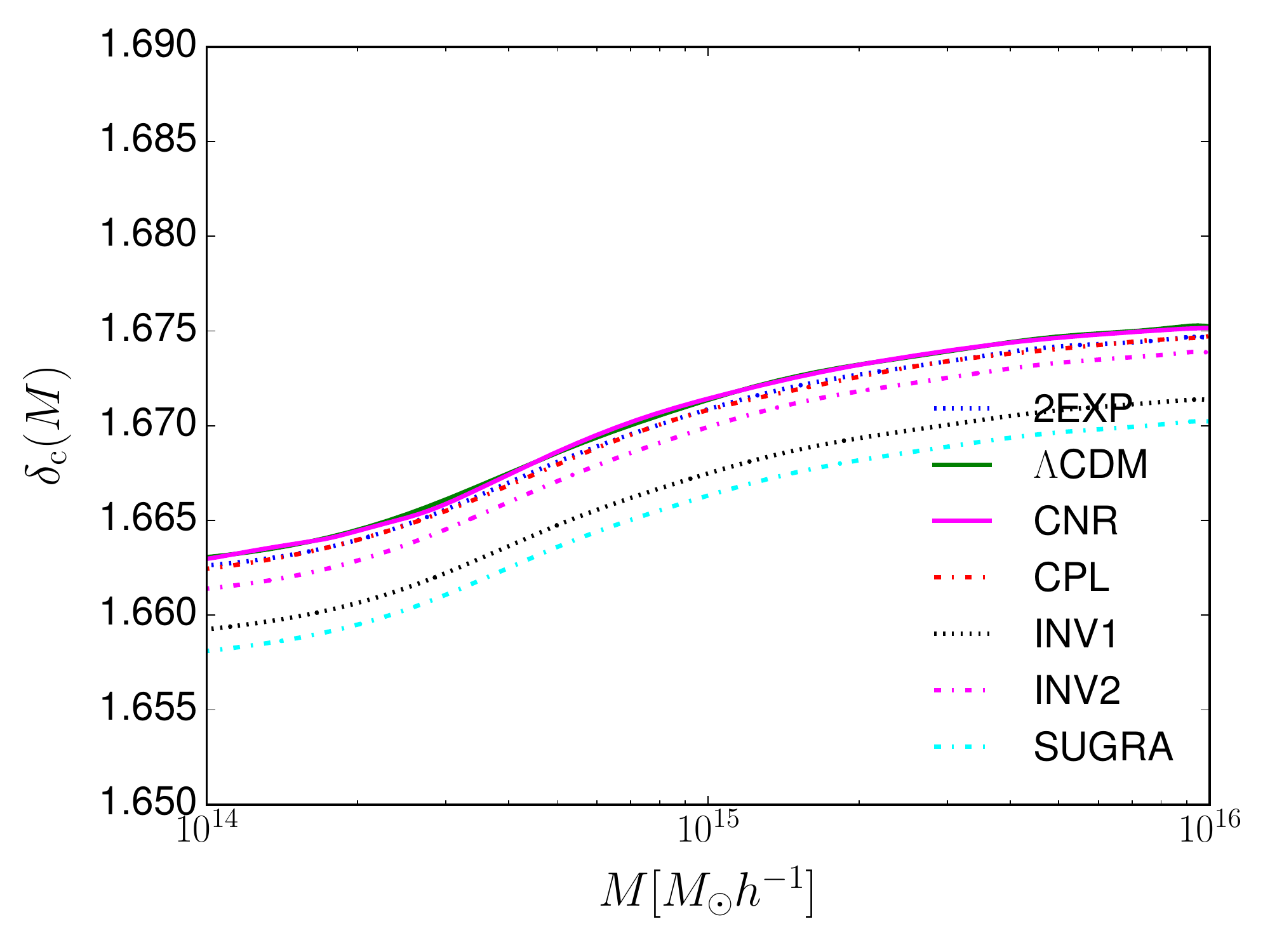}\\
 \includegraphics[width=0.45\textwidth]{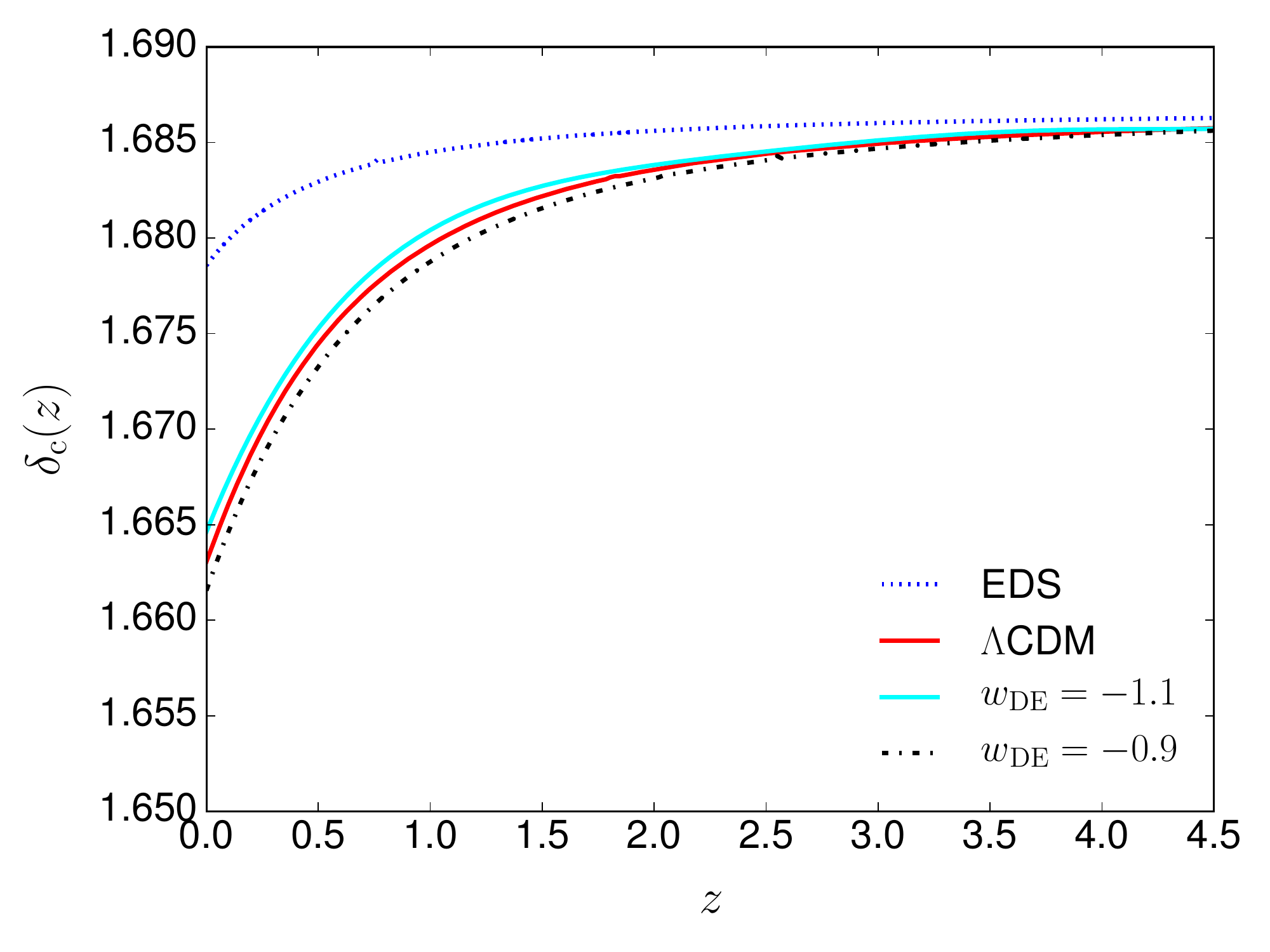}
 \includegraphics[width=0.45\textwidth]{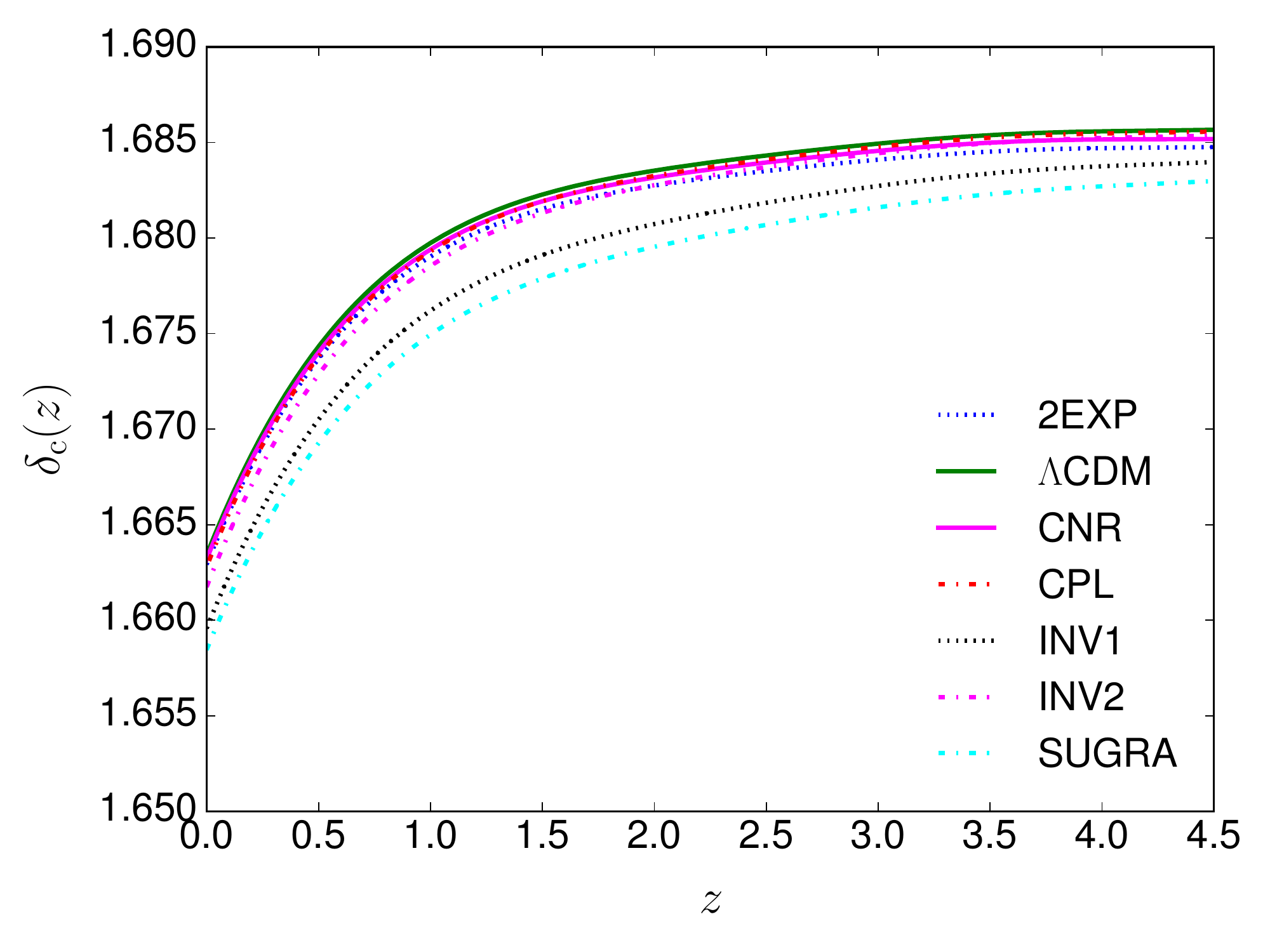}\\
 \includegraphics[width=0.45\textwidth]{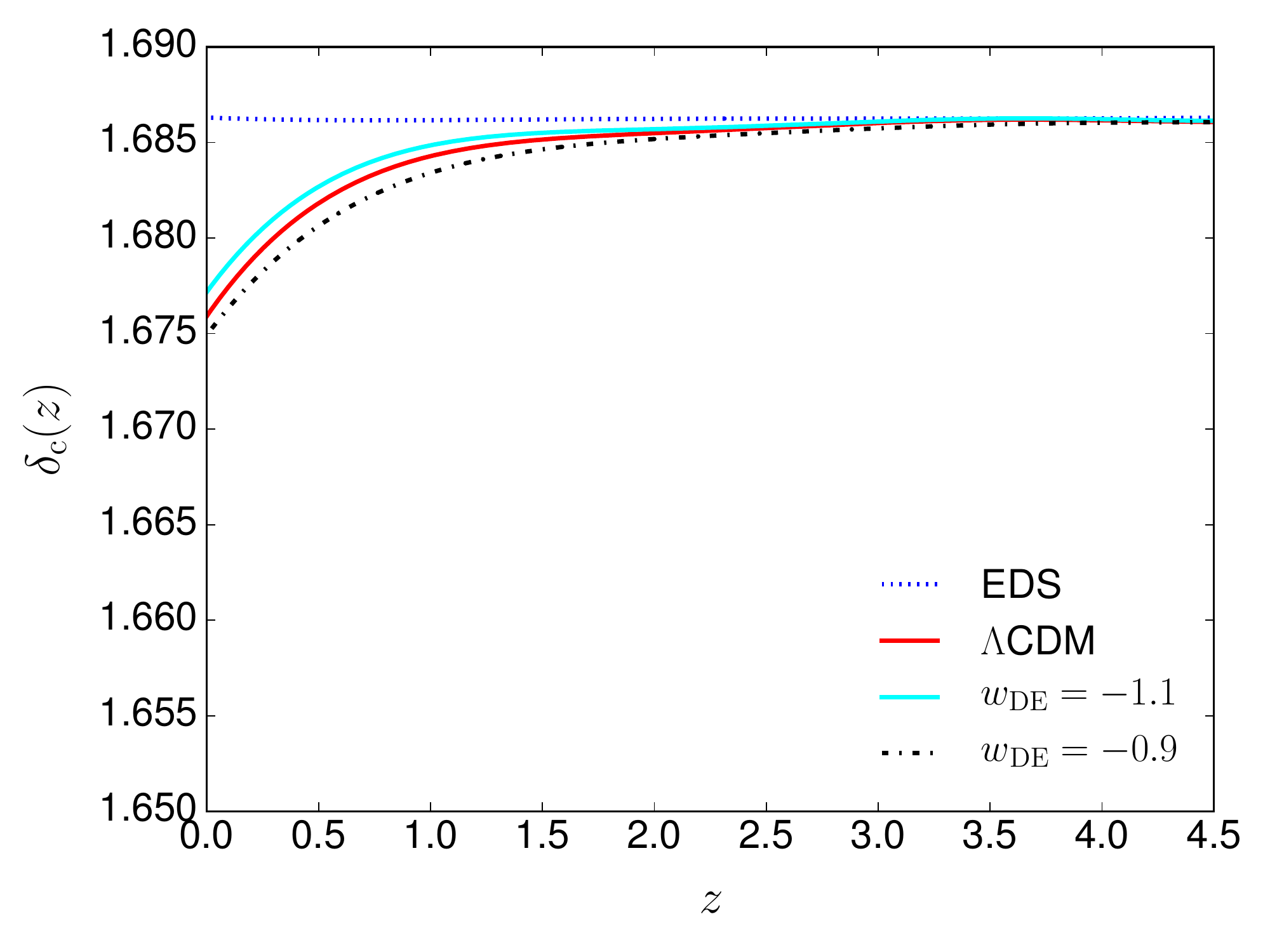}
 \includegraphics[width=0.45\textwidth]{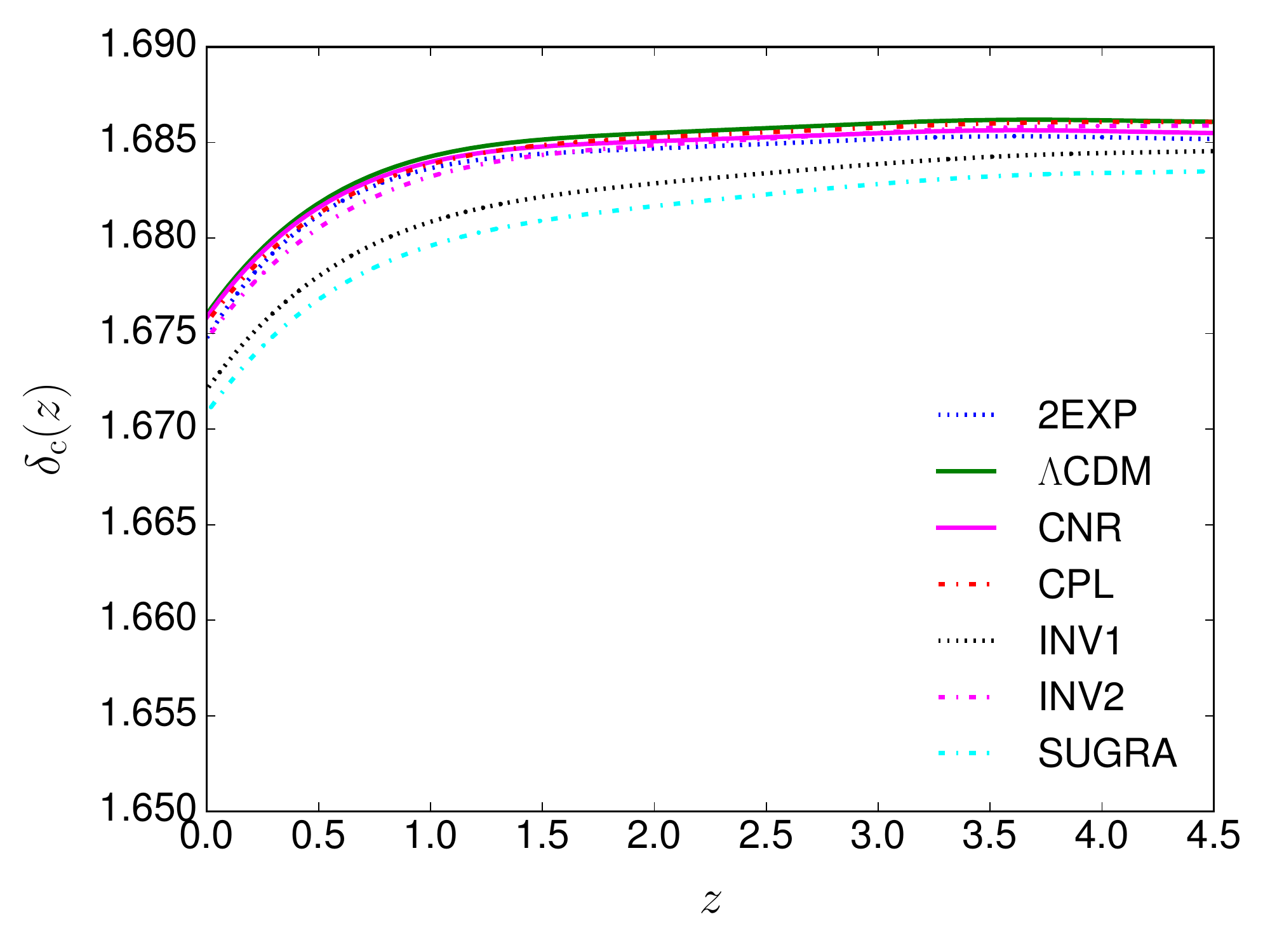}
 \caption{\textit{Upper panels}: effects of the tidal shear on $\delta_{\rm c}$ at $z=0$ for different values of the 
 mass of the collapsing sphere. \textit{Middle panels}: time evolution of $\delta_{\rm c}$ for a mass of 
 $M=10^{14}~M_{\odot}/h$. 
 \textit{Bottom panels}: time evolution of the linear extrapolated overdensity parameter for the standard spherical 
 collapse model (without the inclusion of the tidal shear). Left (right) panels refer to constant (dynamical) 
 equations of state. The red solid line refers to the reference $\Lambda$CDM model. For models with constant 
 equation of state, the blue dotted curve shows an EdS model, while the black dashed-dot (cyan solid) curve shows a 
 quintessence (phantom) model with $w_{\rm DE}=-0.9$ ($w_{\rm DE}=-1.1$). 
 For dynamical dark energy models, the black dotted (magenta dashed-dot) curve represents the INV1 (INV2) 
 model; the blue dashed curve the 2EXP model; the CPL (CNR) model with the red dashed-dot (magenta solid) 
 curve and finally the SUGRA model with the cyan dashed-dot-dotted curve.}
 \label{fig:deltac}
\end{figure*}

In \autoref{fig:deltac} we show our findings for the parameter $\delta_{\rm c}$ in several dark energy models, 
with respect to the $\Lambda$CDM model and to the respective values in absence of tidal shear. We refer the reader to 
the caption for the different colours and line-styles adopted for each model. In the left panels we show models with 
constant equation of state and in the right panels some dynamical models (i.e. with a time-varying equation-of-state 
parameter.)

\begin{figure*}
 \centering
 \includegraphics[width=0.45\textwidth]{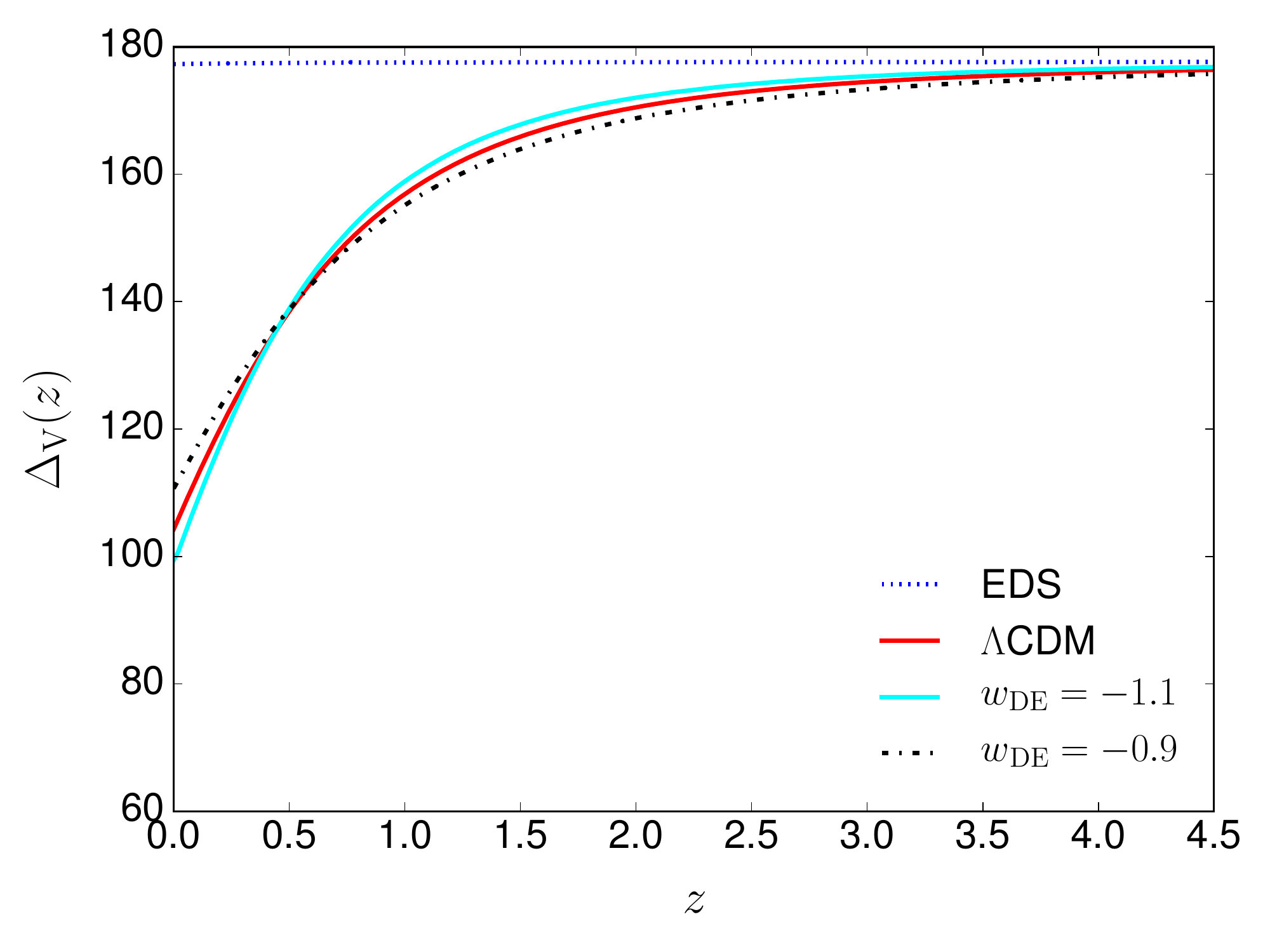}
 \includegraphics[width=0.45\textwidth]{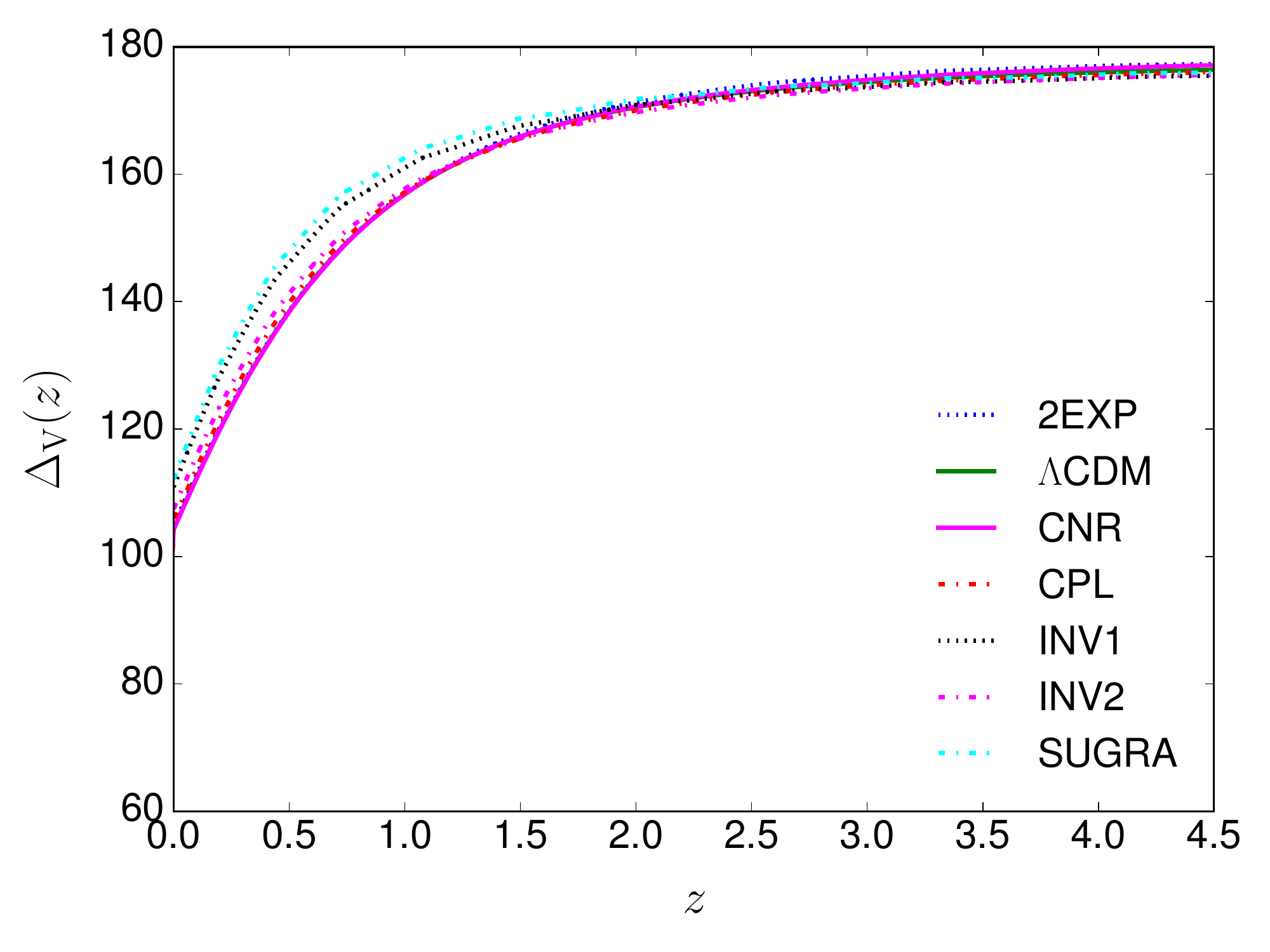}\\
 \includegraphics[width=0.45\textwidth]{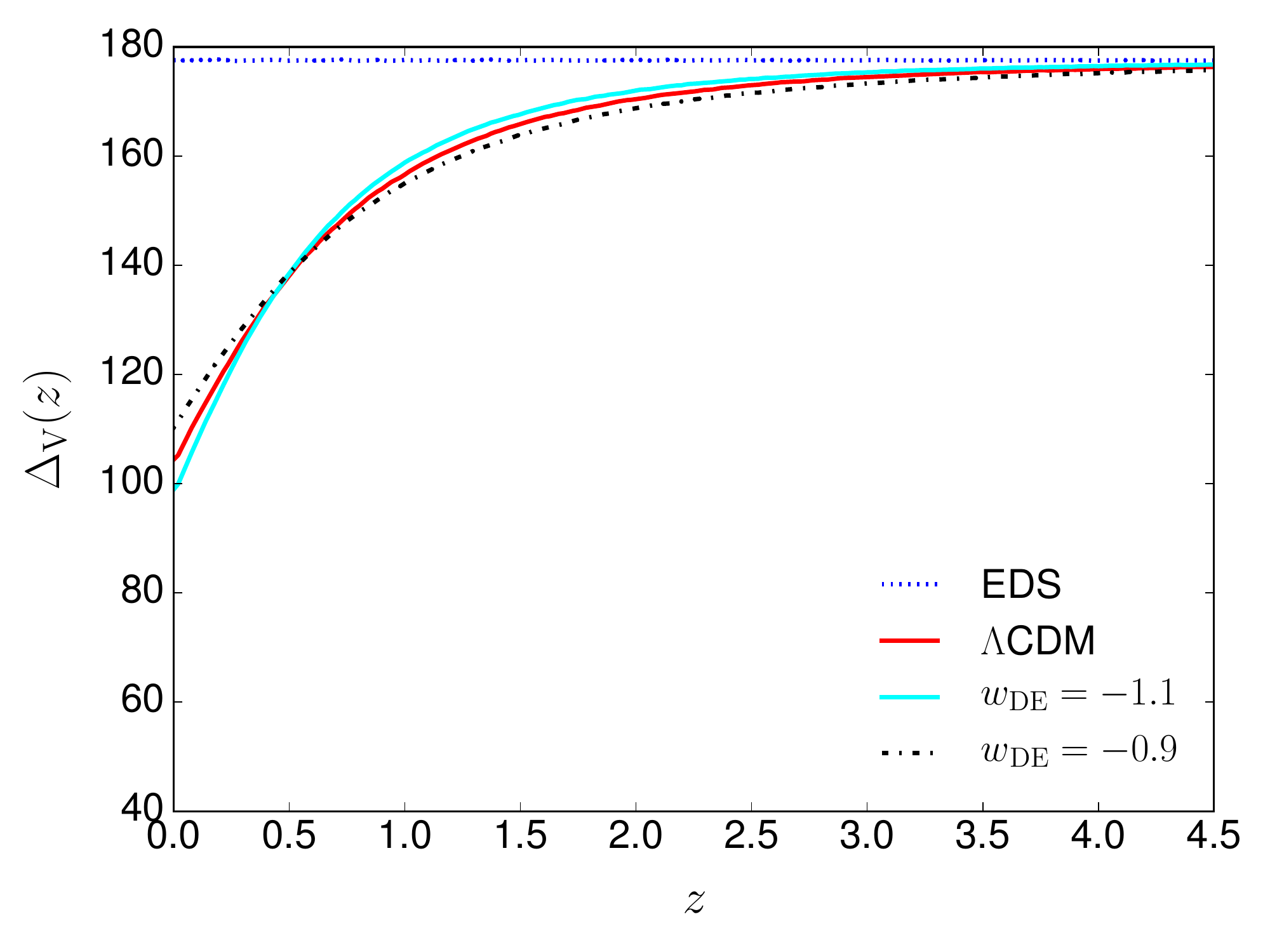}
 \includegraphics[width=0.45\textwidth]{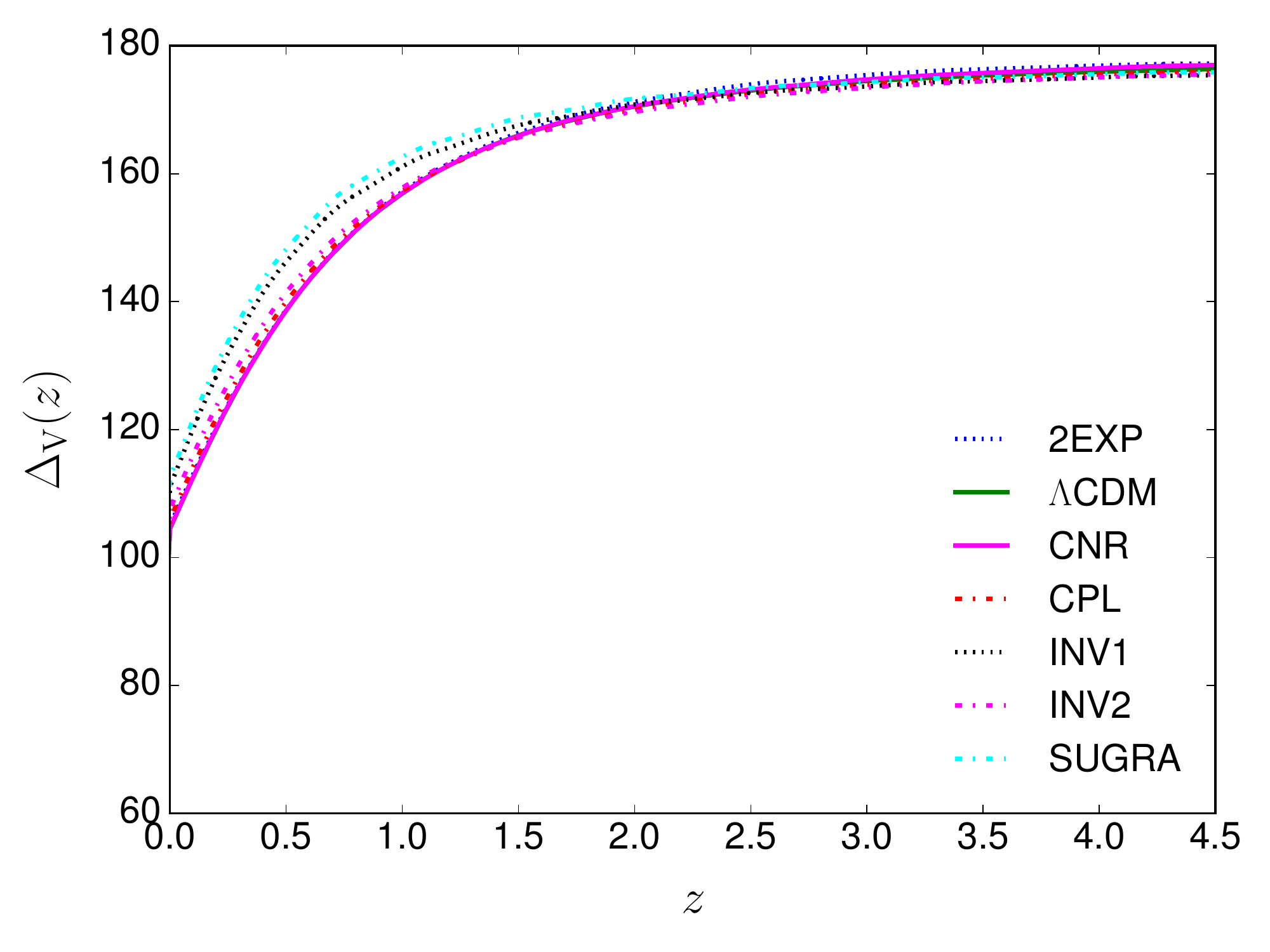}
 \caption{\textit{Upper panels}: time evolution of $\Delta_{\rm V}$ for a mass of $M=10^{14}~M_{\odot}/h$. 
 \textit{Bottom panels}: time evolution of the virial overdensity for the standard spherical collapse model. 
 Left (right) panels refer to constant (dynamical) equations of state. Line-styles and colours are as in 
 \autoref{fig:deltac}.}
 \label{fig:DeltaV}
\end{figure*}

From a first qualitative analysis, results are as expected. Tidal shear favours the collapse and the linearly 
extrapolated overdensity parameter $\delta_{\rm c}$ is smaller than in the spherically symmetric case with no external 
tidal shear (compare the upper panels with values at $z=0$ in the bottom panels). The linear overdensity obviously 
depends on the halo mass now; stronger effects take place at low masses, at high masses the effect is negligible and 
the result converges to the standard spherically symmetric solution. This is particularly evident for the EdS model 
(blue dotted curve). 
Note however that differences from the standard case are quite small, below the 1\% level. 
In the middle panel we fix the mass of the collapsing object at $M=10^{14}~M_{\odot}/h$, so to amplify the effect of 
the tidal shear, and we study the time evolution of the parameter $\delta_{\rm c}$. It is illuminating to compare it 
with the time evolution of the spherically symmetric case (bottom panel) and despite the results are not new since 
already derived and discussed previously in \cite{Pace2010}, we report them once again for clarity. 
First of all notice that due to the tidal shear, for the EdS model $\delta_{\rm c}$ becomes time-dependent. Effects of 
the introduction of the ellipticity are more pronounced at low redshifts and they become negligible at high redshift, 
where the new solution converges to the standard value. Similar results, both qualitatively and quantitatively are 
obtained for generic dark energy models. All the models analysed show lower values for $\delta_{\rm c}$, especially at 
the lower end of the mass interval considered. At high masses values converge to the spherical case. Effects of the 
tidal shear are most evident at low redshifts and negligible at high redshifts. For $z\gtrsim 3$, the tidal shear 
contribution is totally negligible. 
Also for dynamical dark energy models, deviations from the standard case are below 1\%.

In \autoref{fig:DeltaV} we present results for the virial overdensity parameter $\Delta_{\rm V}$. Interestingly, this 
quantity is insensitive to the introduction of the tidal shear and its time evolution is practically identical to what 
observed for the standard spherically symmetric case. This implies that for the virial overdensity, the solution of 
the standard theoretical model is an excellent approximation also for the case including the tidal shear. The reason 
why the results of the two approaches are identical, is due to the fact that the non-linear overdensity at turn-around, 
$\zeta$, is insensitive to the tidal shear. Also note that in general, differences between the dark energy models and 
the $\Lambda$CDM model are very limited.

Our results for $\delta_{\rm c}$ are qualitatively similar to the works of \cite{DelPopolo2013a} and 
\cite{DelPopolo2013b}, albeit with some important differences and, as discussed before, with our formalism we can not 
do a quantitative comparison. First of all, $\delta_{\rm c}$ shows a mass dependence similar to the works mentioned. 
Effects of the modified collapse increase with decreasing mass and at the very high mass tail these modifications 
become negligible. Regarding the time dependence, also in our case larger modifications take place at low redshifts 
and at high enough $z$, the spherical case is an excellent approximation. In \cite{DelPopolo2013a} and 
\cite{DelPopolo2013b}, the authors showed an increase in $\delta_{\rm c}$ rather than a decrease. However, in their 
heuristic model the dominant term was given by the rotation tensor, hence we would expect a slow-down of the collapse. 
It would be therefore interesting to find an approach, similar to what we did here, to take into account also the 
rotation term and then compare the two different approaches. The situation is completely different for the 
virial overdensity $\Delta_{\rm V}$: in our case it is totally independent of the tidal shear, hinting towards the 
hypothesis that probably the rotation is more important or that it is sensitive to the particular prescription adopted. 
Since the virial overdensity $\Delta_{\rm V}$ is largely independent of the tidal shear, it is interesting to examine 
why this happens. First of all it is useful to notice that our values for the tidal shear are much smaller than the 
ones used by Del Popolo and collaborators. To show this, it is sufficient to evaluate the relative strength of the 
shear term with respect to the Poisson term. In other words we are giving an estimate of the parameter $\alpha$ used in 
\mbox{\cite{DelPopolo2013a,DelPopolo2013b}}. We find that in this work, $\alpha\simeq 10^{-8}$, while in previous works 
it was of the order of few per mill for an object of $10^{14}~\text{M}_{\odot}/h$.

To have a physical insight of this, we recall the definition of the virial overdensity. Note that we assume it with 
respect to the critical density, but the same result would apply if we would define it with respect to the background 
density. The virial overdensity is defined as
\begin{equation}
 \Delta_{\rm V}=\zeta\left(\frac{x}{y}\right)^3\;,
\end{equation}
where $\zeta=\delta(a_{\rm ta})+1$ is the overdensity at turn-around, $x=a/a_{\rm ta}$ is the scale factor normalised 
at the turn-around and finally $y=R_{\rm vir}/R_{\rm ta}$ is the virial radius normalized to the turn-around radius. 
As $\sigma^2$ enhances the collapse, $\zeta$ is smaller than the perfectly spherically symmetric case, as we need a 
smaller initial overdensity $\delta_{\rm i}$ to reach the collapse at $z_{\rm c}$. But $\zeta$ is evaluated only in 
the mildly non-linear regime, therefore it is only slightly smaller and the relative contribution of the $\sigma^2$ 
term compared to the Poisson term (the $\alpha$ coefficient) is to be about a few per mill at turn-around, in perfect 
quantitative agreement with our findings about the change of $\Delta_{\rm V}$.

On the other hand $z_{\rm ta}$ is slightly larger, but the effect is really small. 
The virialisation condition leading to $y=R_{\rm vir}/R_{\rm ta}$ does not directly depend on $\sigma^2$, but only 
indirectly via $z_{\rm ta}$. By Taylor expanding $\Delta_{\rm V}$ around the spherically symmetric case ($\sigma^2=0$), 
we have the following relations (the index 0 refers to the absence of shear):
\begin{equation}
 \Delta_{\rm V} = \Delta_{\rm V,0}\left(1 + \frac{\delta\zeta}{\zeta_0} + 
                 3\frac{\delta{z_{\rm ta}}}{1+z_{\rm ta,0}} - 3 \frac{\delta y}{y_0}\right)\;,
\end{equation}
where, for a $\Lambda$CDM model at $z_{\rm c}=0$, we have:
\begin{equation}
 \frac{\delta\zeta}{\zeta_0} \simeq -0.015 \;, \quad
 \frac{\delta{z_{\rm ta}}}{1+z_{\rm ta,0}} \simeq 0.0014 \; \quad 
 \frac{\delta y}{y_0} \simeq -0.0047\;. \nonumber
\end{equation}
It is therefore clear that albeit extremely small, the dominant contribution is due to the change in $\zeta$, making 
as expected the virial overdensity only slightly smaller than in the spherical case.

It is also interesting to make a more direct comparison with the ellipsoidal collapse. One of the goals of this work is 
to establish whether a Press-Schechter formulation of the mass function with the corrections induced on $\delta_{\rm 
c}$ by the tidal shear tensor could give predictions closer to a Sheth-Tormen formulation with the standard 
$\delta_{\rm c}$ values. According to \cite{Bond1996a}, the collapse time depends on the ellipticity $e$ and 
prolaticity $p$ and the dependence of the collapse threshold of an ellipsoidal region can be well approximated by the 
solution of \citep{Sheth2001}
\begin{equation}
 \frac{\delta_{\rm ec}}{\delta_{\rm sc}}=1+\beta\left[5(e^2\pm p^2)\frac{\delta_{\rm ec}^2}{\delta_{\rm sc}^2}
 \right]^{\gamma}\;,
\end{equation}
where $\delta_{\rm ec}$ and $\delta_{\rm sc}$ are the values of the critical overdensity for the ellipsoidal and 
spherical case, respectively and $\beta$ and $\gamma$ are parameters to be fitted to the results. 
\cite{Doroshkevich1970} and \cite{Sheth2001} found that
\begin{equation}
 \delta_{\rm ec} = \delta_{\rm sc}\left[1+\beta\left(\frac{\sigma(M)^2}{\delta_{\rm sc}}\right)^{\gamma}\right]\;,
\end{equation}
with $\beta=0.47$ and $\gamma=0.615$. With $\sigma(M)$ of the order unity, $\delta_{\rm ec}$ is about 25\% - 30\% 
bigger than $\delta_{\rm sc}$. We can therefore conclude that the tidal shear will have small effects on the mass 
function, as shown later in \autoref{fig:DEmf}.

\section{Mass function}\label{sec:MF}
The halo mass function describes the differential abundance of objects with mass $M$ at redshift $z$. Working within 
the theory of Gaussian random fields, the main ingredient is the comparison of fluctuations of the linearly evolved 
density field with $\delta_\text{c}$. Objects exceeding $\delta_\text{c}$ on a certain scale $R(M)$ are then counted 
as clusters. The fluctuations of the density field are described by the variance $\sigma_R$ of the underlying random 
field filtered with a top-hat having a certain scale. \citet{Press1974} showed that the mass function (PS) has the 
form
\begin{equation}
 n(M,z) = \frac{2\rho_0}{\pi M} \frac{\delta_\text{c}(z)}{D_+\sigma_R}\left|\frac{\partial\ln\sigma_R}{\partial M}
 \right|\exp\left(-\frac{\delta_\text{c}^2(z)}{2D_+^2(z)\sigma_R^2}\right)\;,
\end{equation}
where the growth factor $D_+$ accounts for the linear evolution. More elaborate forms of the mass function, fitting 
numerical $N$-body simulations better, are given in \citet{Sheth1999} or \citet{Jenkins2001}. 
The important functional form for our purpose is however given by the term
\begin{equation}
 \frac{\delta_\text{c}(z)}{D_+\sigma_R}\exp\left(-\frac{\delta_\text{c}^2(z)}{2D_+^2(z)\sigma_R^2}\right),
\end{equation}
where we replace
\begin{equation}\label{eq:mfshear}
 \delta_\text{c}(z) \to \bar{\delta}_\text{c}(M,z)\;,
\end{equation}
i.e. we insert the effective $\delta_\text{c}$. This has important consequences: Firstly, $\delta_\text{c}$ changes 
with the mass which will lead to a different form of the mass function. Furthermore, the shear causes $\delta_\text{c}$ 
to be smaller than without shear as it only supports the collapse. Due to the functional form of the mass function we 
therefore expect more massive haloes in the mass regime where the exponential factor dominates the linear one. On 
smaller scales, however, the linear term will dominate, thus causing the mass function to tend to smaller values. The 
reason for this behaviour is that small haloes can form more massive haloes more easily, thus yielding fewer smaller 
objects. 
Finally the time dependence of the shear is different from the linear growth of $\sigma_\text{R}$, we thus expect 
different impacts of the shear on different redshifts which in principle can make $\Lambda$CDM and $w$CDM models 
degenerate.

We now want to infer the influence of the tidal shear on the mass function for the several dark energy models analysed 
in this work. To do so, we evaluate the cumulative comoving number density of objects above a given mass at $z=0$. For 
all the models we assume $\sigma_8=0.776$. This is done not to introduce volume and normalization effects that would 
mask the contribution of the tidal shear, that, as we will see, amounts to few percent in a $\Lambda$CDM model.

\begin{figure*}
 \includegraphics[width=0.45\textwidth]{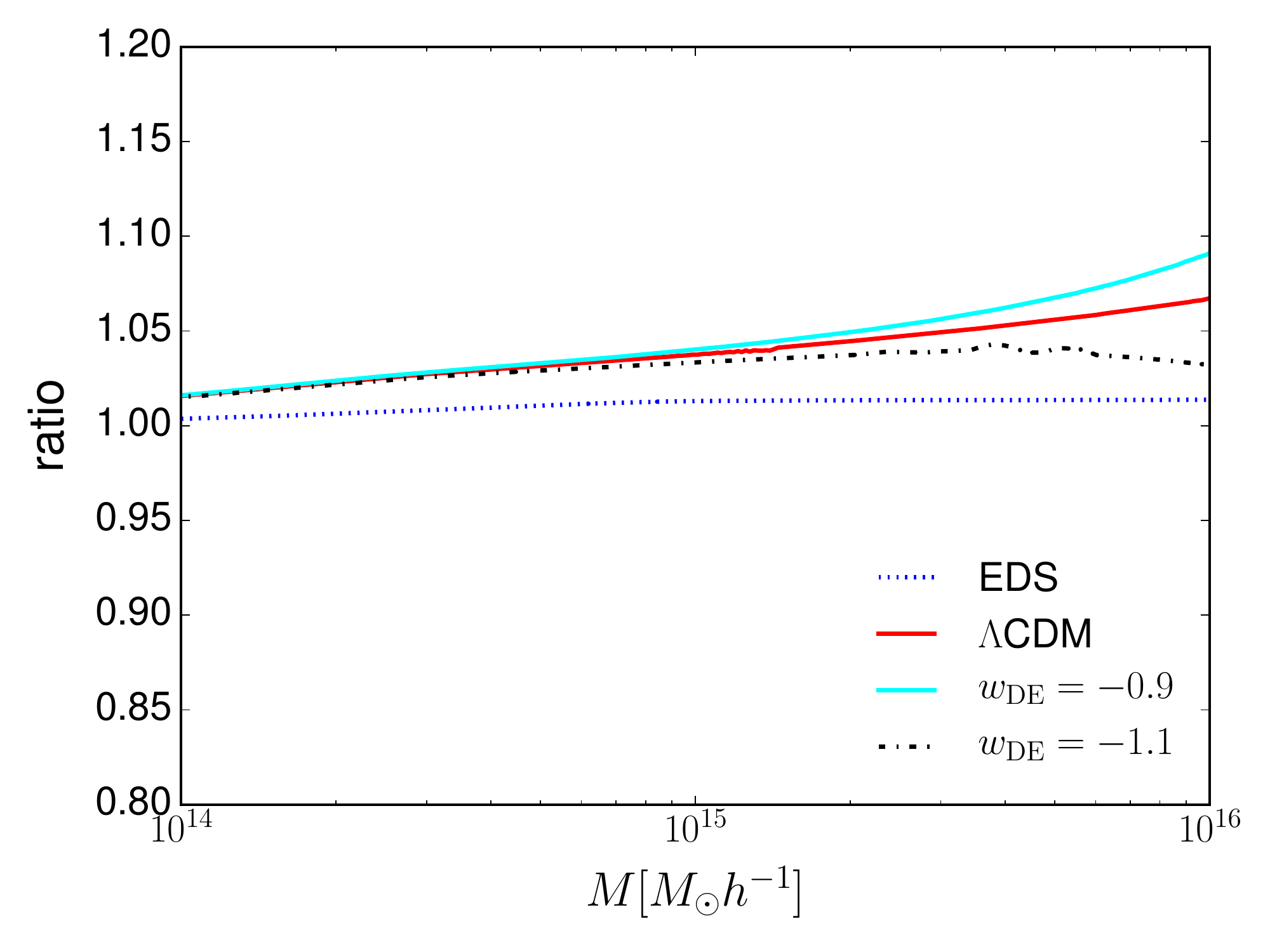}
 \includegraphics[width=0.45\textwidth]{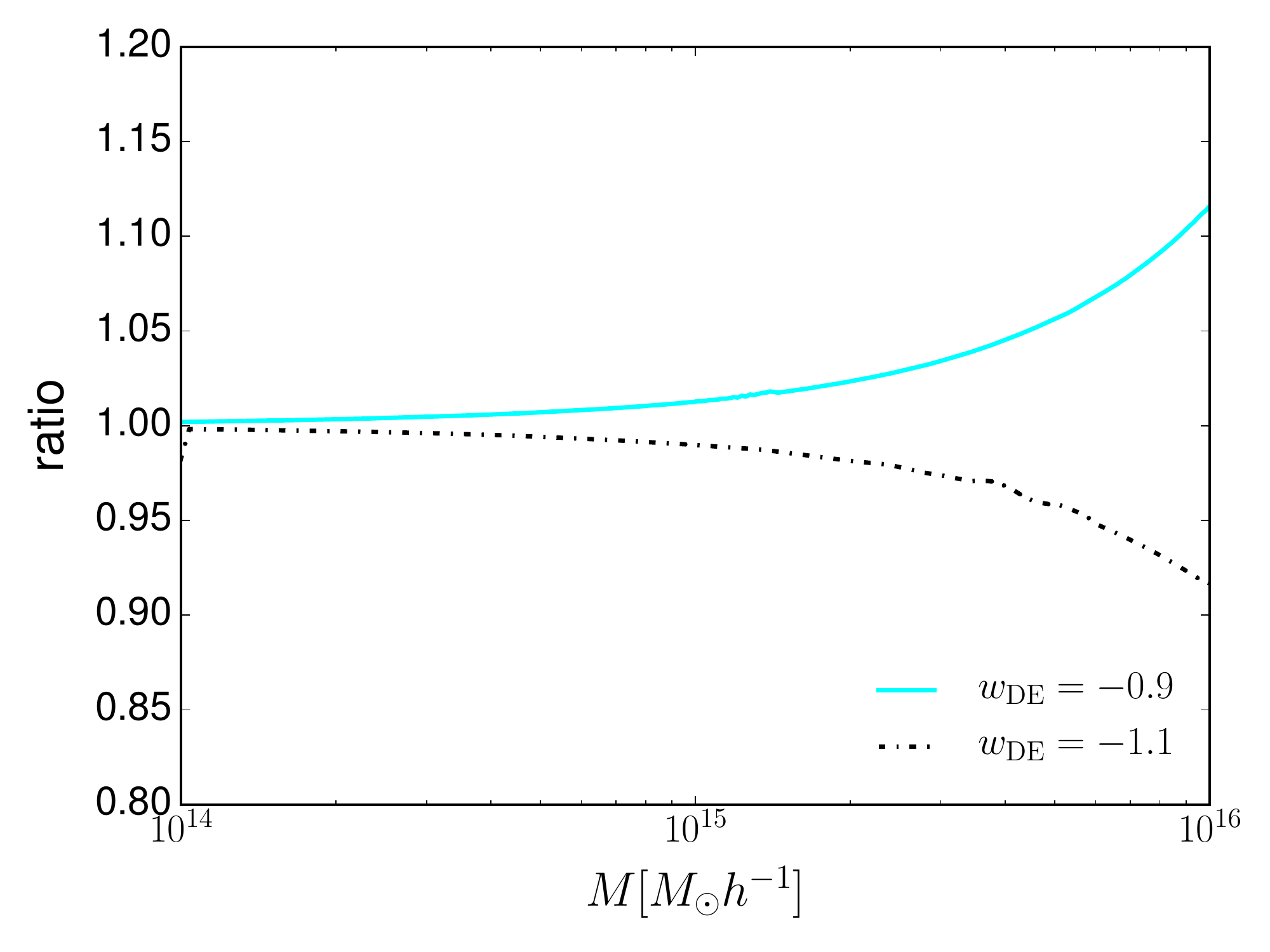}\\
 \includegraphics[width=0.45\textwidth]{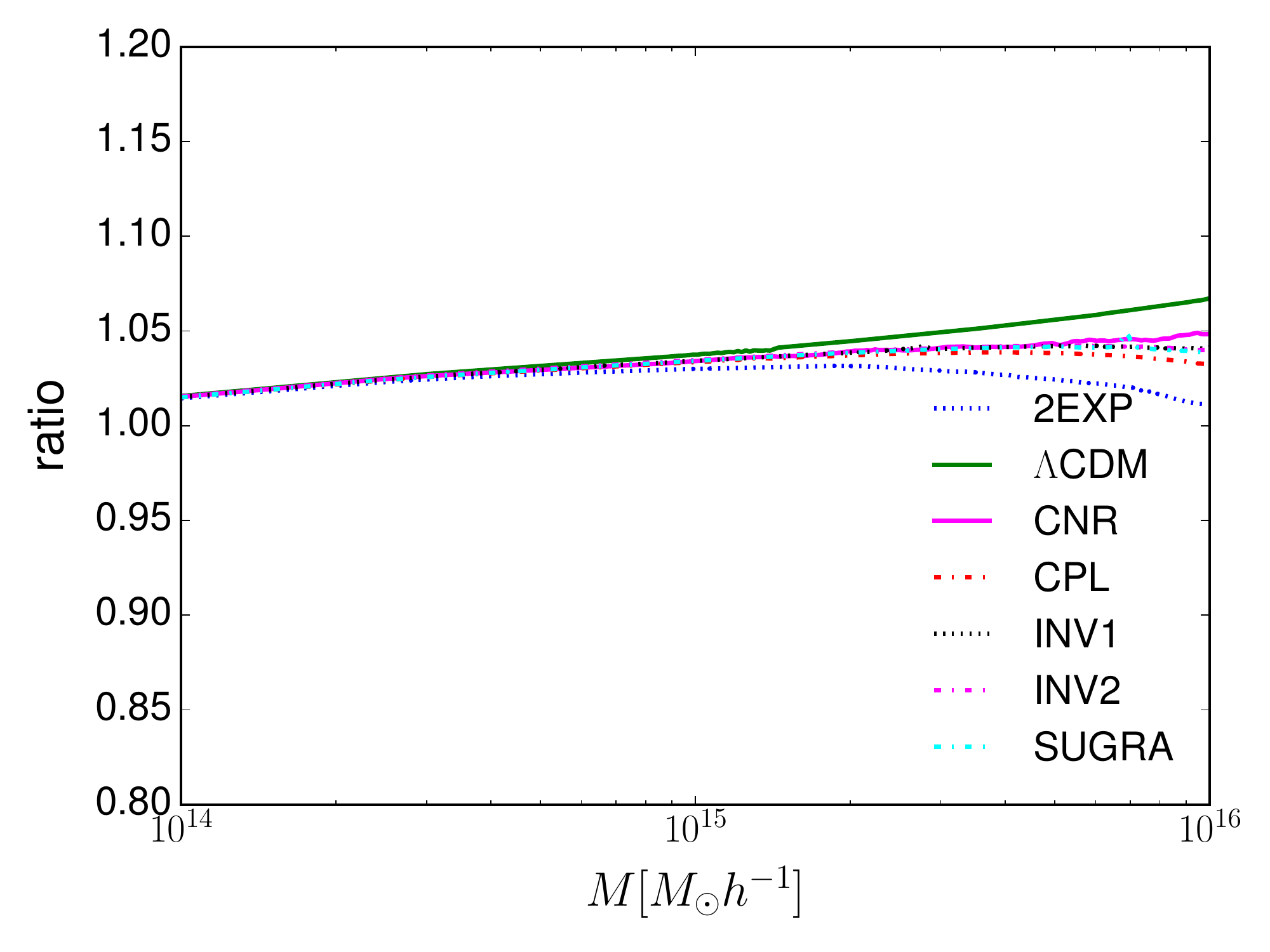}
 \includegraphics[width=0.45\textwidth]{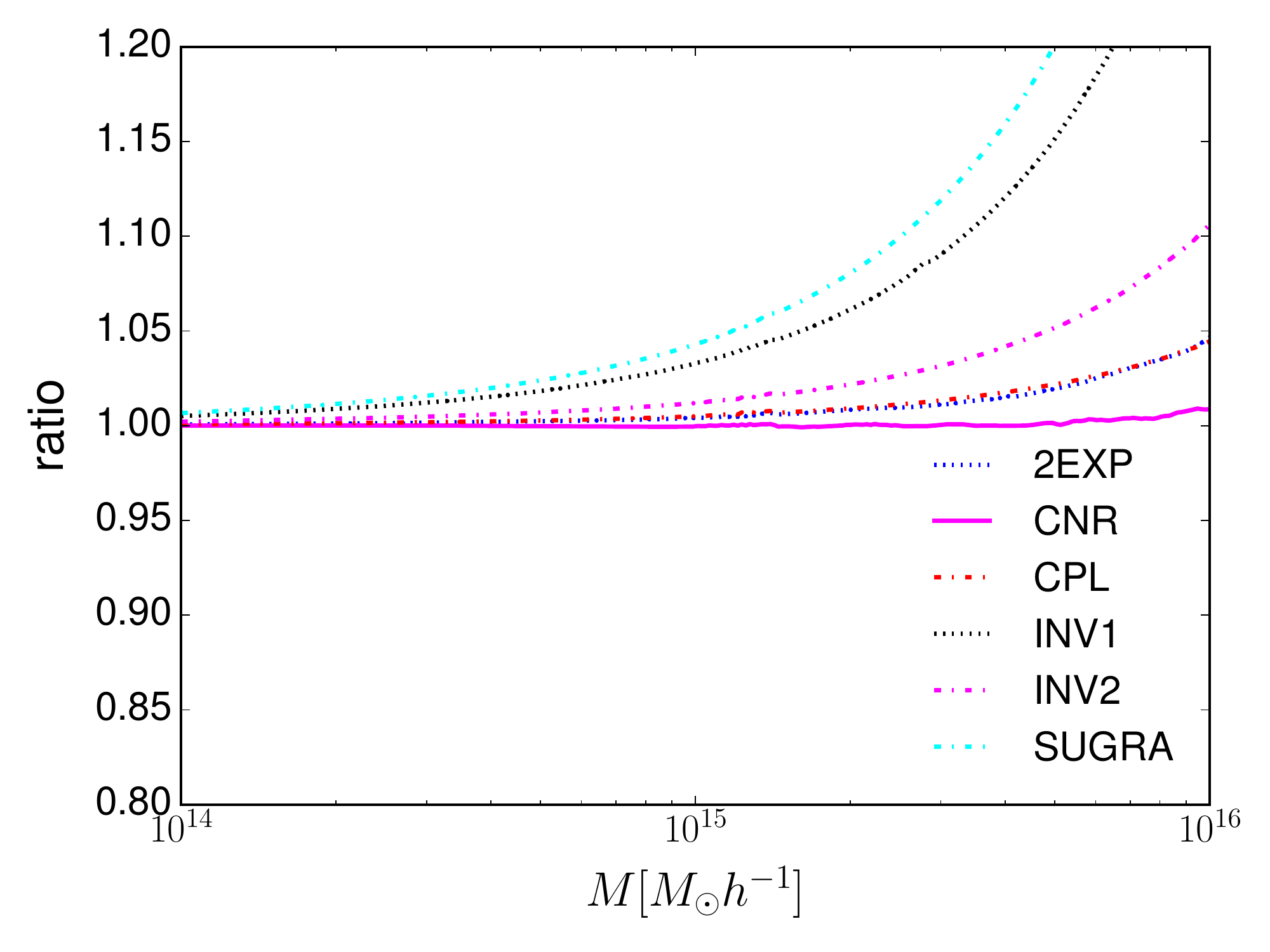}
 \caption{Ratio of the cumulative comoving number density of objects above mass $M$ evaluated at $z=0$. 
 \textit{Left column}: ratio between the expected number counts of the models with and without tidal shear 
 contribution. 
 \textit{Right column}: ration between the dark energy and the $\Lambda$CDM model taking into account the effects of 
 the tidal shear. Upper panels show results for the EdS model and dark energy models with constant equation of state 
 ($w_{\rm DE}=-0.9$ and $w_{\rm DE}=-1.1$). Lower panels show results for dynamical dark energy models. 
 Line-styles and colours are as in \autoref{fig:deltac}.}
 \label{fig:DEmf}
\end{figure*}

We present our results in \autoref{fig:DEmf}, where we show the ratio between the dark energy and the reference 
$\Lambda$CDM model. In the left panels we show the ratio between the models with and without tidal shear field while in 
the right panels we show the ratio between the dark energy and the $\Lambda$CDM model with tidal shear field.

By inspecting the left panels, we realize that, as expected, tidal shear has a modest contribution, usually growing 
with increasing mass. The effect is of the order of few percent at the lower limit of applicability of our formalism 
($M\approx 10^{14}~M_{\odot}/h$) and it increases up to 10\% for a model with constant $w_{\rm DE}=-0.9$ at very high 
masses ($M\approx 10^{16}~M_{\odot}/h$). The model being least affected is the EdS, somehow in agreement with what 
found for the spherical collapse parameter $\delta_{\rm c}$. 
Interestingly, the SUGRA model shows an increase with mass up to $M\simeq 10^{15}~M_{\odot}/h$ and a slow decrease 
to bring the model with tidal shear close to the standard one. 
Note that however differences are never bigger than about 3\% for this model. Also note that, except for the model with 
constant $w_{\rm DE}=-0.9$, all the other models show an effect less pronounced in the high mass tail than the 
$\Lambda$CDM model and all the models, except for the EdS one, are identical to the $\Lambda$CDM model up to masses of 
$\simeq 10^{15}~M_{\odot}/h$.

In the right panels we show the ratio between the dark energy models and the $\Lambda$CDM one, both with the effects 
of the tidal shear field included. Results are both qualitatively and quantitatively as expected. For models with 
constant equation of state, the quintessence (phantom) model predicts more (less) objects with respect to the 
$\Lambda$CDM model and differences grow increasing the halo mass. All the dynamical dark energy models are in the 
quintessence regime and we see, as expected, more objects than the $\Lambda$CDM one. The CNR model behaves essentially 
as the $\Lambda$CDM model and the models CPL and 2EXP are practically indistinguishable and predict about 5\% more 
objects than the $\Lambda$CDM one. Major differences arise for the SUGRA and the INV1 model.

The mass function described in \cite{Sheth1999} was introduced, as said above, to have a better match with $N$-body 
simulations. To do so, the authors incorporated the effect of shear in their calculations within the formalism of the 
ellipsoidal collapse model. The main quantity characterising the mass function is still the ratio $\delta_{\rm c}/D_+$ 
and effects due to the ellipsoidal collapse are incorporated directly in functional form of the mass function. 
We can therefore try to answer the following question: Will the Press-Schechter mass function approximate better the 
Sheth-Tormen mass function by using relation~(\ref{eq:mfshear})? The idea behind that is in principle incorporating the 
tidal shear effects into the linear overdensity parameter and making it mass-dependent could compensate the necessity 
of 
modifying the functional form of the mass function. However, as one can see already in \autoref{fig:deltac}, the 
influence on the mass function will only be a few percent. Accordingly it will of course improve the agreement between 
the Sheth Tormen mass function and the PS mass function, nonetheless this improvement is rather marginal with respect 
to 
the differences of the two mass functions at the high mass end. This can also be seen from the elliptical collapse 
model where the effective influence on $\delta_\text{c}$ is much larger than in our case leading to a big change in the 
number counts due to the exponential tail for massive objects including the collapse threshold (see \autoref{sec:spc} 
for a more detailed discussion).

\begin{figure*}
 \includegraphics[width=0.45\textwidth]{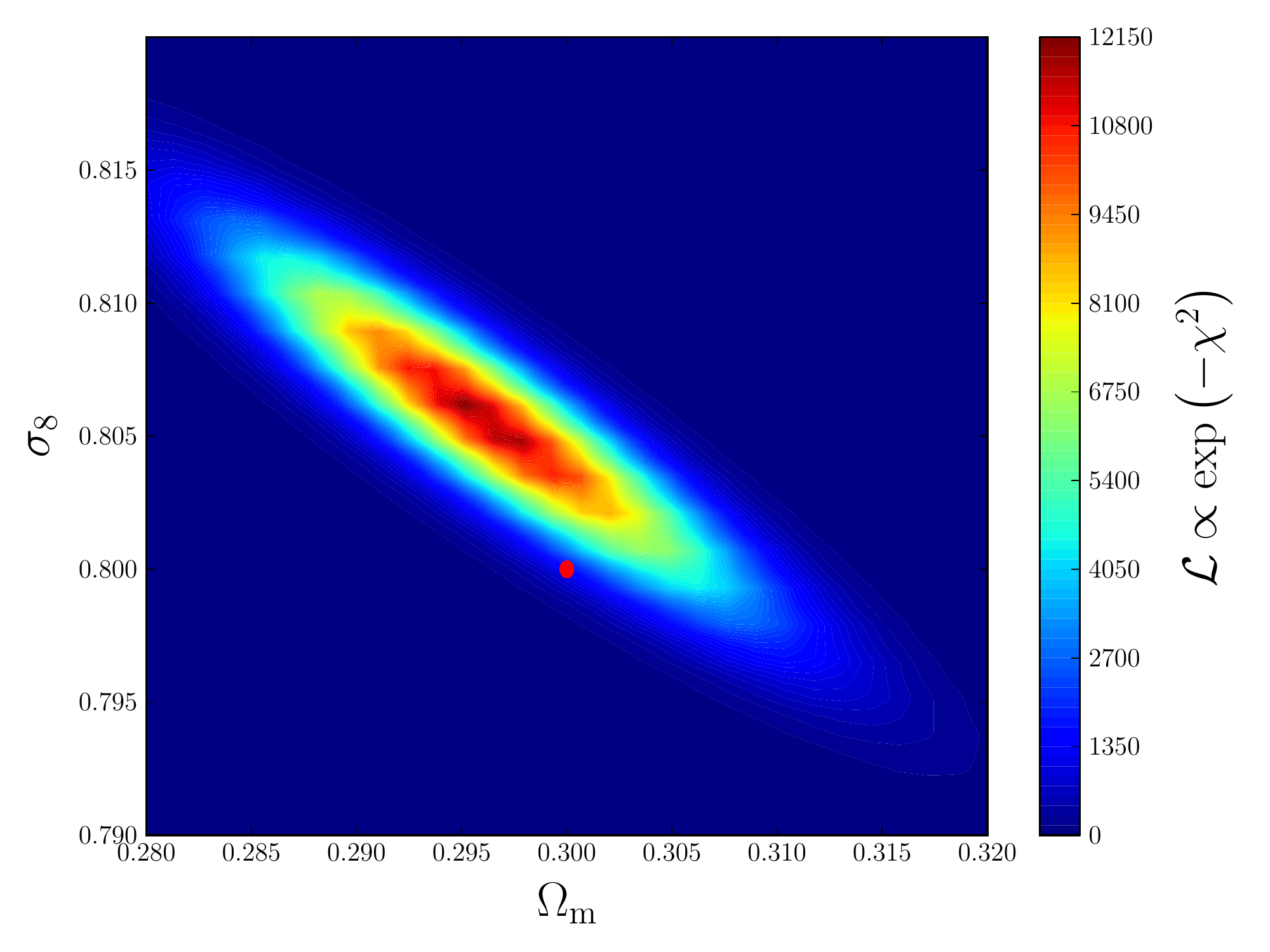}
 \includegraphics[width=0.45\textwidth]{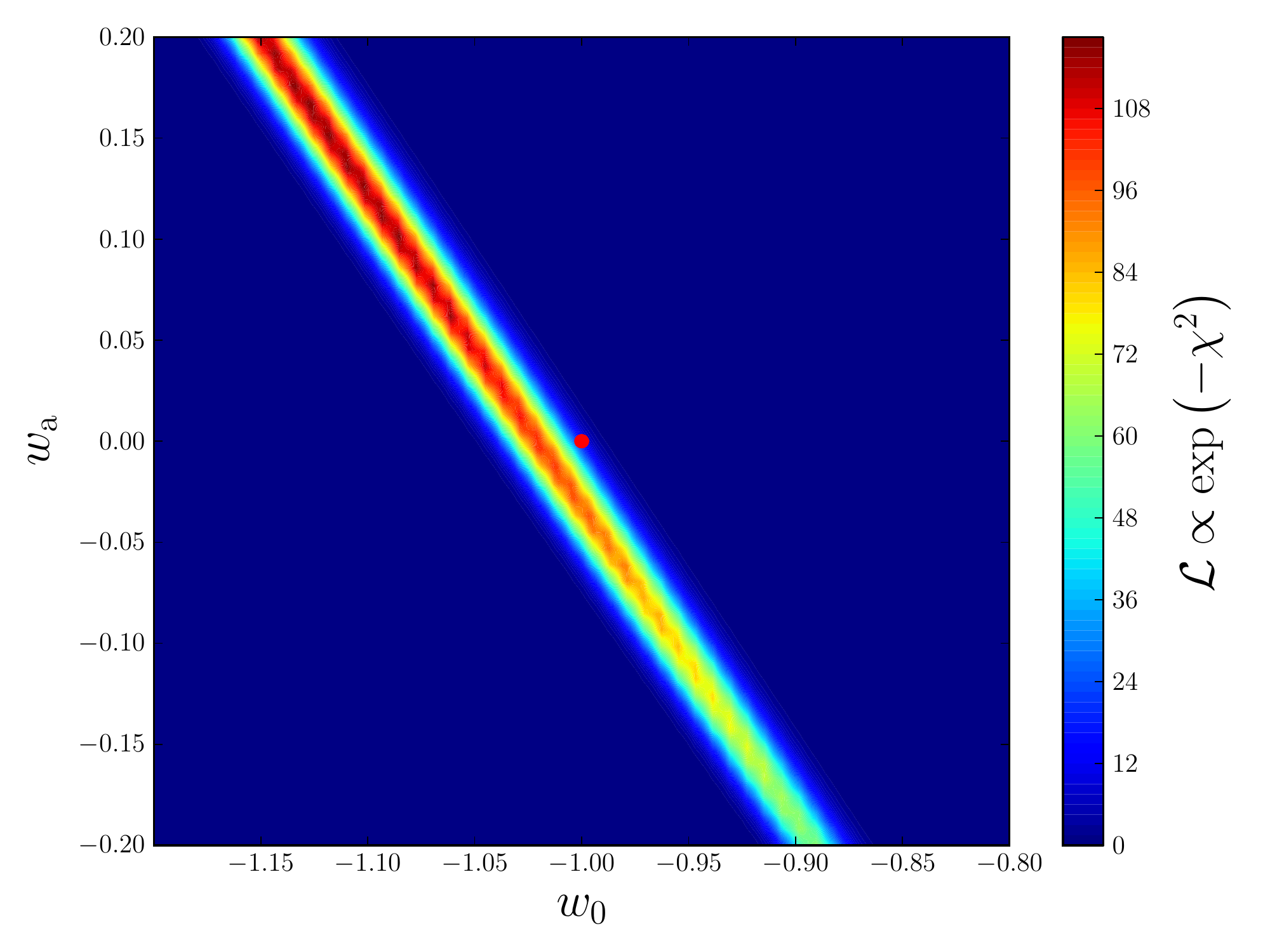}
 \caption{Bias on cosmological parameters for redshift cluster counts. Parameters not shown in the respective plot are 
 fixed to their fiducial values. The red dot marks the fiducial cosmology with shear.}
 \label{Fig:8}
\end{figure*}

\section{Cluster Counts}\label{sec:clustercounts}
From the mass function, cluster counts can be calculated, which can then be compared to observational data. Using 
Sunyaev-Zel'dovich cluster surveys \citep{Sunyaev1980b}, the number of objects exceeding a mass $M_\text{min}$ 
in a redshift bin $z_i$ is given by \citep{Majumdar2004}
\begin{equation}
 N(z_i)\equiv N_i = 4\pi f_\text{sky}\int_{z_i-\Delta z_i/2}^{z_i+\Delta z_i/2}
 \text{d}z\frac{\text{d}V}{\text{d}z}\int_{M_\text{min}(z)}^\infty\text{d}M\; n(M,z)\;,
\end{equation}
where $f_\text{sky}$ is the fraction of the sky. $M_\text{min}$ has a redshift dependence included. Assuming a 
Gaussian likelihood, the log-likelihood is given by $\chi^2$:
\begin{equation}
2{\chi^2} \equiv L = \sum_{i} \frac{(N_i-\langle N_i\rangle)^2}{N_i}\;,
\end{equation}
where we sum over all redshift bins and Poisson errors are assumed. $\langle N_i\rangle$ is the model dependent 
expected number of objects in the $i$-th bin, while $N_i$ describes the data. We chose redshift bins with 
$\Delta z=0.02$ ranging from $z_\text{min} = 0.01$ to $z_\text{max} = 2$. For simplicity we assume 
$M_\text{min} = 10^{14}h^{-1} M_{\odot}$ to be redshift independent. The mock data is sampled from a Poisson 
distribution with mean $N_i$ evaluated at the fiducial cosmology with $\Omega_\text{m0}=0.3$, $\Omega_{\Lambda} = 0.7$, 
$\sigma_8=0.8$, $w_0 = -1$ and $w_\text{a}=1$ and including shear effects in the mass function, cf. 
Eq.~(\ref{eq:mfshear}). A cosmological model without shear effects is fitted to this data, leading to biases in the 
cosmological parameters.

In \autoref{Fig:8} we show the resulting biases in parameter space. The red dot marks the fiducial cosmology at which 
the data was sampled from a mass function including tidal shear via $\delta_\text{c}$. In contrast the black dot marks 
the best fit value of cosmological models without shear effects acting on $\delta_\text{c}$. Ignoring shear effects 
accordingly leads to wrong cosmological parameters, which are shifted by $\sim 1\sigma$ with respect to the true 
values for both $(\Omega_\text{m},\sigma_8)$ and $(w_0,w_\text{a})$.

\section{Conclusion}\label{sec:concl}
In this work we investigated the influence of external tidal shear effects on the spherical collapse model using first 
order Lagrangian perturbation theory. The shear is evaluated directly from the statistics of the underlying density 
field in which the halo forms and therefore it does not need any further assumptions. Clearly, we cannot include 
rotational effects with our formalism, as the rotation vanishes identically for a potential flow, however for the 
scales investigated the assumption of linear growth is still valid, implying that, even if initial rotation was present 
(which would require third order Lagrangian perturbation theory) it would decay as the halo forms. In this sense our 
Ansatz for the shear is self-consistent.

In contrast, shear effects become more important for lower redshifts, as also the structures in the vicinity of the 
collapsing objects grow, thus increasing the curvature of the potential. We summarize our findings as follows:

\begin{enumerate}

\item External tidal shear supports the spherical collapse, which can be understood by noticing that virialized objects 
form in overdense regions in the first place. The effect is largest at small masses and low redshifts.

\item The effect on the important collapse parameter $\delta_\text{c}$ is of the percent level for both $\Lambda$CDM 
and more general dark energy models. 
Furthermore the influence on the virial overdensity $\Delta_\text{V}$ is very small and it is nearly indistinguishable 
from a collapse without external shear. The reason for this is mainly that the virial overdensity is basically 
evaluated using the time at turn-around. At this time the evolution is only in the mildly non-linear regime, therefore 
shear effects are not important.

\item Gaining a mass dependence due to our formalism, the influence on the mass function is two-fold. At lower masses 
the linear term dominates the exponential, suppressing the occurrence of lighter objects. Furthermore the influence of 
$\delta_\text{c}$ is largest at high masses, as the exponential tail dominates there. However, the effect of shear on 
$\delta_\text{c}$ becomes smaller for higher masses. The mentioned effects leads to a change of the mass function of 
roughly $2\%$.

\item The mass dependence translates into differences also in the cumulative number counts. Tidal shear affects number counts of 
massive halos of only few percent when compared to the corresponding model without it. When compared to the 
$\Lambda$CDM model with tidal shear, results are qualitatively and quantitatively the same as without tidal shear.

\item Neglecting the shear in the estimation of cosmological parameters using number counts, e.g. in redshift space, 
can lead to $1\sigma$ biases on cosmological parameters such as $\Omega_{\rm m}$, $\sigma_8$, $w_0$ and $w_\text{a}$.

\item The bias in the cosmological parameters is such that the inferred $\sigma_8$ ($\Omega_{\rm m}$) is higher (lower) than without and a $\Lambda$CDM model results into a dynamical phantom model for $a\approx 1$. The increase in $\sigma_8$ is in the right direction to at least alleviate the tension between the power spectrum normalization at late and early times, even if the amount is not sufficient. Remember however, that we neglected the rotation contribution and this could either balance or strengthen the shear contribution. Also the resulting phantom model is in agreement with SNIa observations, but at this stage we cannot draw any firm conclusion.

\item Previous works on the extended spherical collapse model introduced the effect of shear and rotation with 
heuristically motivated models \citep{DelPopolo2013a}. In this model, both shear and rotation are combined into a 
single term that depends on mass and result into a modification of the Poisson term, but the rotation term has a 
predominant role with respect to the shear. While a direct comparison cannot be made, our approach has some points in 
common and some major differences. While both approaches lead to a mass dependent spherical collapse, we find that 
effects of the shear are at percent level, contrary to what found in previous works. This leads to the 
question of the importance of the rotation and of its effective modelization.

\end{enumerate}

\section*{Acknowledgements}
Parts of the basic calculations (background cosmological model or halo model implementations) have been 
performed using the C++ implementations developed and maintained in the research group of Matthias Bartelmann.
RR acknowledges funding by the graduate college Astrophysics
of cosmological probes of gravity by Landesgraduiertenakademie
Baden-W\"urttemberg. FP acknowledges useful discussions about the work at the UK Cosmo meeting in Sussex.
\bibliographystyle{mnras}
\bibliography{old_MasterBib.bib,MyBiB.bib}

\begin{thebibliography}{}
\makeatletter
\relax
\def\mn@urlcharsother{\let\do\@makeother \do\$\do\&\do\#\do\^\do\_\do\%\do\~}
\def\mn@doi{\begingroup\mn@urlcharsother \@ifnextchar [ {\mn@doi@}
  {\mn@doi@[]}}
\def\mn@doi@[#1]#2{\def\@tempa{#1}\ifx\@tempa\@empty \href
  {http://dx.doi.org/#2} {doi:#2}\else \href {http://dx.doi.org/#2} {#1}\fi
  \endgroup}
\def\mn@eprint#1#2{\mn@eprint@#1:#2::\@nil}
\def\mn@eprint@arXiv#1{\href {http://arxiv.org/abs/#1} {{\tt arXiv:#1}}}
\def\mn@eprint@dblp#1{\href {http://dblp.uni-trier.de/rec/bibtex/#1.xml}
  {dblp:#1}}
\def\mn@eprint@#1:#2:#3:#4\@nil{\def\@tempa {#1}\def\@tempb {#2}\def\@tempc
  {#3}\ifx \@tempc \@empty \let \@tempc \@tempb \let \@tempb \@tempa \fi \ifx
  \@tempb \@empty \def\@tempb {arXiv}\fi \@ifundefined
  {mn@eprint@\@tempb}{\@tempb:\@tempc}{\expandafter \expandafter \csname
  mn@eprint@\@tempb\endcsname \expandafter{\@tempc}}}

\bibitem[\protect\citeauthoryear{{Abramo}, {Batista}, {Liberato}  \&
  {Rosenfeld}}{{Abramo} et~al.}{2007}]{Abramo2007}
{Abramo} L.~R.,  {Batista} R.~C.,  {Liberato} L.,   {Rosenfeld} R.,  2007,
  \mn@doi [Journal of Cosmology and Astro-Particle Physics]
  {10.1088/1475-7516/2007/11/012}, \href
  {http://adsabs.harvard.edu/abs/2007JCAP...11..012A} {11, 12}

\bibitem[\protect\citeauthoryear{{Abramo}, {Batista}  \& {Rosenfeld}}{{Abramo}
  et~al.}{2009}]{Abramo2009a}
{Abramo} L.~R.,  {Batista} R.~C.,   {Rosenfeld} R.,  2009, \mn@doi [Journal of
  Cosmology and Astro-Particle Physics] {10.1088/1475-7516/2009/07/040}, \href
  {http://adsabs.harvard.edu/abs/2009JCAP...07..040A} {7, 40}

\bibitem[\protect\citeauthoryear{Angrick \& Bartelmann}{Angrick \&
  Bartelmann}{2009}]{Angrick2009}
Angrick C.,  Bartelmann M.,  2009, Astronomy \& Astrophysics, 494, 461

\bibitem[\protect\citeauthoryear{Angrick \& Bartelmann}{Angrick \&
  Bartelmann}{2010}]{Angrick2010}
Angrick C.,  Bartelmann M.,  2010, Astronomy \& Astrophysics, 518, A38

\bibitem[\protect\citeauthoryear{{Avila-Reese}, {Firmani}  \&
  {Hern{\'a}ndez}}{{Avila-Reese} et~al.}{1998}]{AvilaReese1998}
{Avila-Reese} V.,  {Firmani} C.,   {Hern{\'a}ndez} X.,  1998, \mn@doi [\apj]
  {10.1086/306136}, \href {http://adsabs.harvard.edu/abs/1998ApJ...505...37A}
  {505, 37}

\bibitem[\protect\citeauthoryear{{Barreiro}, {Copeland}  \& {Nunes}}{{Barreiro}
  et~al.}{2000}]{Barreiro2000}
{Barreiro} T.,  {Copeland} E.~J.,   {Nunes} N.~J.,  2000, \mn@doi [\prd]
  {10.1103/PhysRevD.61.127301}, \href
  {http://adsabs.harvard.edu/abs/2000PhRvD..61l7301B} {61, 127301}

\bibitem[\protect\citeauthoryear{{Bernardeau}}{{Bernardeau}}{1994}]{Bernardeau1994}
{Bernardeau} F.,  1994, \mn@doi [\apj] {10.1086/174620}, \href
  {http://adsabs.harvard.edu/abs/1994ApJ...433....1B} {433, 1}

\bibitem[\protect\citeauthoryear{{Bertschinger}}{{Bertschinger}}{1985}]{Bertschinger1985}
{Bertschinger} E.,  1985, \mn@doi [\apjs] {10.1086/191028}, \href
  {http://adsabs.harvard.edu/abs/1985ApJS...58...39B} {58, 39}

\bibitem[\protect\citeauthoryear{{Bond} \& {Myers}}{{Bond} \&
  {Myers}}{1996}]{Bond1996a}
{Bond} J.~R.,  {Myers} S.~T.,  1996, \mn@doi [\apjs] {10.1086/192267}, \href
  {http://adsabs.harvard.edu/abs/1996ApJS..103....1B} {103, 1}

\bibitem[\protect\citeauthoryear{{Chevallier} \& {Polarski}}{{Chevallier} \&
  {Polarski}}{2001}]{Chevallier2001}
{Chevallier} M.,  {Polarski} D.,  2001, \mn@doi [International Journal of
  Modern Physics D] {10.1142/S0218271801000822}, \href
  {http://adsabs.harvard.edu/abs/2001IJMPD..10..213C} {10, 213}

\bibitem[\protect\citeauthoryear{{Cole} et~al.,}{{Cole}
  et~al.}{2005}]{Cole2005}
{Cole} S.,  et~al., 2005, \mn@doi [\mnras] {10.1111/j.1365-2966.2005.09318.x},
  \href {http://adsabs.harvard.edu/abs/2005MNRAS.362..505C} {362, 505}

\bibitem[\protect\citeauthoryear{{Copeland}, {Nunes}  \& {Rosati}}{{Copeland}
  et~al.}{2000}]{Copeland2000}
{Copeland} E.~J.,  {Nunes} N.~J.,   {Rosati} F.,  2000, \mn@doi [\prd]
  {10.1103/PhysRevD.62.123503}, \href
  {http://adsabs.harvard.edu/abs/2000PhRvD..62l3503C} {62, 123503}

\bibitem[\protect\citeauthoryear{{Copeland}, {Sami}  \& {Tsujikawa}}{{Copeland}
  et~al.}{2006}]{Copeland2006}
{Copeland} E.~J.,  {Sami} M.,   {Tsujikawa} S.,  2006, \mn@doi [International
  Journal of Modern Physics D] {10.1142/S021827180600942X}, \href
  {http://adsabs.harvard.edu/abs/2006IJMPD..15.1753C} {15, 1753}

\bibitem[\protect\citeauthoryear{{Corasaniti}}{{Corasaniti}}{2004}]{Corasaniti2004}
{Corasaniti} P.~S.,  2004, PhD thesis, University of Sussex, \url
  {http://de.arxiv.org/pdf/astro-ph/0401517}

\bibitem[\protect\citeauthoryear{{Corasaniti} \& {Copeland}}{{Corasaniti} \&
  {Copeland}}{2003}]{Corasaniti2003}
{Corasaniti} P.~S.,  {Copeland} E.~J.,  2003, \mn@doi [\prd]
  {10.1103/PhysRevD.67.063521}, \href
  {http://adsabs.harvard.edu/abs/2003PhRvD..67f3521C} {67, 063521}

\bibitem[\protect\citeauthoryear{{Del Popolo}, {Pace}  \& {Lima}}{{Del Popolo}
  et~al.}{2013a}]{DelPopolo2013a}
{Del Popolo} A.,  {Pace} F.,   {Lima} J.~A.~S.,  2013a, \mn@doi [International
  Journal of Modern Physics D] {10.1142/S0218271813500387}, \href
  {http://adsabs.harvard.edu/abs/2013IJMPD..2250038D} {22, 50038}

\bibitem[\protect\citeauthoryear{{Del Popolo}, {Pace}  \& {Lima}}{{Del Popolo}
  et~al.}{2013b}]{DelPopolo2013b}
{Del Popolo} A.,  {Pace} F.,   {Lima} J.~A.~S.,  2013b, \mn@doi [\mnras]
  {10.1093/mnras/sts669}, \href
  {http://adsabs.harvard.edu/abs/2013MNRAS.430..628D} {430, 628}

\bibitem[\protect\citeauthoryear{{Diego} \& {Majumdar}}{{Diego} \&
  {Majumdar}}{2004}]{Diego2004}
{Diego} J.~M.,  {Majumdar} S.,  2004, \mn@doi [\mnras]
  {10.1111/j.1365-2966.2004.07989.x}, \href
  {http://adsabs.harvard.edu/abs/2004MNRAS.352..993D} {352, 993}

\bibitem[\protect\citeauthoryear{{Doroshkevich}}{{Doroshkevich}}{1970}]{Doroshkevich1970}
{Doroshkevich} A.~G.,  1970, Astrofizika, \href
  {http://adsabs.harvard.edu/abs/1970Afz.....6..581D} {6, 581}

\bibitem[\protect\citeauthoryear{{Fang} \& {Haiman}}{{Fang} \&
  {Haiman}}{2007}]{Fang2007}
{Fang} W.,  {Haiman} Z.,  2007, \mn@doi [\prd] {10.1103/PhysRevD.75.043010},
  \href {http://adsabs.harvard.edu/abs/2007PhRvD..75d3010F} {75, 043010}

\bibitem[\protect\citeauthoryear{{Fillmore} \& {Goldreich}}{{Fillmore} \&
  {Goldreich}}{1984}]{Fillmore1984}
{Fillmore} J.~A.,  {Goldreich} P.,  1984, \mn@doi [\apj] {10.1086/162070},
  \href {http://adsabs.harvard.edu/abs/1984ApJ...281....1F} {281, 1}

\bibitem[\protect\citeauthoryear{{Gunn} \& {Gott}}{{Gunn} \&
  {Gott}}{1972}]{Gunn1972}
{Gunn} J.~E.,  {Gott} III J.~R.,  1972, \mn@doi [\apj] {10.1086/151605}, \href
  {http://adsabs.harvard.edu/abs/1972ApJ...176....1G} {176, 1}

\bibitem[\protect\citeauthoryear{Heavens \& Sheth}{Heavens \&
  Sheth}{1999}]{Heavens1999}
Heavens A.~F.,  Sheth R.~K.,  1999, Monthly Notices of the Royal Astronomical
  Society, 310, 1062

\bibitem[\protect\citeauthoryear{{Jenkins}, {Frenk}, {White}, {Colberg},
  {Cole}, {Evrard}, {Couchman}  \& {Yoshida}}{{Jenkins}
  et~al.}{2001}]{Jenkins2001}
{Jenkins} A.,  {Frenk} C.~S.,  {White} S.~D.~M.,  {Colberg} J.~M.,  {Cole} S.,
  {Evrard} A.~E.,  {Couchman} H.~M.~P.,   {Yoshida} N.,  2001, \mn@doi [\mnras]
  {10.1046/j.1365-8711.2001.04029.x}, \href
  {http://adsabs.harvard.edu/abs/2001MNRAS.321..372J} {321, 372}

\bibitem[\protect\citeauthoryear{{Komatsu}, {Smith}, {Dunkley}  \&
  {et~al.}}{{Komatsu} et~al.}{2011}]{Komatsu2011}
{Komatsu} E.,  {Smith} K.~M.,  {Dunkley} J.,   {et~al.} 2011, \mn@doi [\apjs]
  {10.1088/0067-0049/192/2/18}, \href
  {http://adsabs.harvard.edu/abs/2011ApJS..192...18K} {192, 18}

\bibitem[\protect\citeauthoryear{Lin \& Kilbinger}{Lin \&
  Kilbinger}{2014}]{Lin2014}
Lin C.-A.,  Kilbinger M.,  2014, Proceedings of the International Astronomical
  Union, 10, 107

\bibitem[\protect\citeauthoryear{{Linder}}{{Linder}}{2003}]{Linder2003}
{Linder} E.~V.,  2003, \mn@doi [Physical Review Letters]
  {10.1103/PhysRevLett.90.091301}, \href
  {http://adsabs.harvard.edu/abs/2003PhRvL..90i1301L} {90, 091301}

\bibitem[\protect\citeauthoryear{{Majumdar}}{{Majumdar}}{2004}]{Majumdar2004}
{Majumdar} S.,  2004, \mn@doi [Pramana] {10.1007/BF02705209}, \href
  {http://adsabs.harvard.edu/abs/2004Prama..63..871M} {63, 871}

\bibitem[\protect\citeauthoryear{{Maturi}, {Angrick}, {Pace}  \&
  {Bartelmann}}{{Maturi} et~al.}{2010}]{Maturi2010}
{Maturi} M.,  {Angrick} C.,  {Pace} F.,   {Bartelmann} M.,  2010, \mn@doi
  [\aap] {10.1051/0004-6361/200912866}, \href
  {http://adsabs.harvard.edu/abs/2010A%26A...519A..23M} {519, A23}

\bibitem[\protect\citeauthoryear{Maturi, Fedeli  \& Moscardini}{Maturi
  et~al.}{2011}]{Maturi2011}
Maturi M.,  Fedeli C.,   Moscardini L.,  2011, Monthly Notices of the Royal
  Astronomical Society, 416, 2527

\bibitem[\protect\citeauthoryear{{Mota} \& {van de Bruck}}{{Mota} \& {van de
  Bruck}}{2004}]{Mota2004}
{Mota} D.~F.,  {van de Bruck} C.,  2004, \mn@doi [\aap]
  {10.1051/0004-6361:20041090}, \href
  {http://adsabs.harvard.edu/abs/2004A%26A...421...71M} {421, 71}

\bibitem[\protect\citeauthoryear{{Ohta}, {Kayo}  \& {Taruya}}{{Ohta}
  et~al.}{2003}]{Ohta2003}
{Ohta} Y.,  {Kayo} I.,   {Taruya} A.,  2003, \mn@doi [\apj] {10.1086/374375},
  \href {http://adsabs.harvard.edu/abs/2003ApJ...589....1O} {589, 1}

\bibitem[\protect\citeauthoryear{{Ohta}, {Kayo}  \& {Taruya}}{{Ohta}
  et~al.}{2004}]{Ohta2004}
{Ohta} Y.,  {Kayo} I.,   {Taruya} A.,  2004, \mn@doi [\apj] {10.1086/420762},
  \href {http://adsabs.harvard.edu/abs/2004ApJ...608..647O} {608, 647}

\bibitem[\protect\citeauthoryear{{Pace}, {Waizmann}  \& {Bartelmann}}{{Pace}
  et~al.}{2010}]{Pace2010}
{Pace} F.,  {Waizmann} J.-C.,   {Bartelmann} M.,  2010, \mn@doi [\mnras]
  {10.1111/j.1365-2966.2010.16841.x}, \href
  {http://adsabs.harvard.edu/abs/2010MNRAS.406.1865P} {406, 1865}

\bibitem[\protect\citeauthoryear{{Pace}, {Fedeli}, {Moscardini}  \&
  {Bartelmann}}{{Pace} et~al.}{2012}]{Pace2012}
{Pace} F.,  {Fedeli} C.,  {Moscardini} L.,   {Bartelmann} M.,  2012, \mn@doi
  [\mnras] {10.1111/j.1365-2966.2012.20692.x}, \href
  {http://adsabs.harvard.edu/abs/2012MNRAS.422.1186P} {422, 1186}

\bibitem[\protect\citeauthoryear{{Pace}, {Moscardini}, {Crittenden},
  {Bartelmann}  \& {Pettorino}}{{Pace} et~al.}{2014a}]{Pace2014}
{Pace} F.,  {Moscardini} L.,  {Crittenden} R.,  {Bartelmann} M.,   {Pettorino}
  V.,  2014a, \mn@doi [\mnras] {10.1093/mnras/stt1907}, \href
  {http://adsabs.harvard.edu/abs/2014MNRAS.437..547P} {437, 547}

\bibitem[\protect\citeauthoryear{{Pace}, {Batista}  \& {Del Popolo}}{{Pace}
  et~al.}{2014b}]{Pace2014b}
{Pace} F.,  {Batista} R.~C.,   {Del Popolo} A.,  2014b, \mn@doi [\mnras]
  {10.1093/mnras/stu1782}, \href
  {http://adsabs.harvard.edu/abs/2014MNRAS.445..648P} {445, 648}

\bibitem[\protect\citeauthoryear{{Padmanabhan}}{{Padmanabhan}}{1996}]{Padmanabhan1996}
{Padmanabhan} T.,  1996, {Cosmology and Astrophysics through Problems}

\bibitem[\protect\citeauthoryear{{Perlmutter}, {Aldering}, {Goldhaber}  \& {et
  al.}}{{Perlmutter} et~al.}{1999}]{Perlmutter1999}
{Perlmutter} S.,  {Aldering} G.,  {Goldhaber} G.,   {et al.} 1999, \mn@doi
  [\apj] {10.1086/307221}, \href
  {http://adsabs.harvard.edu/abs/1999ApJ...517..565P} {517, 565}

\bibitem[\protect\citeauthoryear{{Planck Collaboration XIII}}{{Planck
  Collaboration XIII}}{2015}]{Planck2015_XIII}
{Planck Collaboration XIII} 2015, ArXiv e-prints, 1502.01589, \href
  {http://adsabs.harvard.edu/abs/2015arXiv150201589P} {}

\bibitem[\protect\citeauthoryear{{Press} \& {Schechter}}{{Press} \&
  {Schechter}}{1974}]{Press1974}
{Press} W.~H.,  {Schechter} P.,  1974, \mn@doi [\apj] {10.1086/152650}, \href
  {http://adsabs.harvard.edu/abs/1974ApJ...187..425P} {187, 425}

\bibitem[\protect\citeauthoryear{Reg\H{o}s \& Szalay}{Reg\H{o}s \&
  Szalay}{1995}]{Regos1995}
Reg\H{o}s E.,  Szalay A.~S.,  1995, Monthly Notices of the Royal Astronomical
  Society, 272, 447

\bibitem[\protect\citeauthoryear{Reischke, Maturi  \& Bartelmann}{Reischke
  et~al.}{2016}]{Reischke2016}
Reischke R.,  Maturi M.,   Bartelmann M.,  2016, Monthly Notices of the Royal
  Astronomical Society, 456, 641

\bibitem[\protect\citeauthoryear{{Riess}, {Filippenko}, {Challis}  \& {et
  al.}}{{Riess} et~al.}{1998}]{Riess1998}
{Riess} A.~G.,  {Filippenko} A.~V.,  {Challis} P.,   {et al.} 1998, \mn@doi
  [\aj] {10.1086/300499}, \href
  {http://adsabs.harvard.edu/abs/1998AJ....116.1009R} {116, 1009}

\bibitem[\protect\citeauthoryear{{Ryden} \& {Gunn}}{{Ryden} \&
  {Gunn}}{1987}]{Ryden1987}
{Ryden} B.~S.,  {Gunn} J.~E.,  1987, \mn@doi [\apj] {10.1086/165349}, \href
  {http://adsabs.harvard.edu/abs/1987ApJ...318...15R} {318, 15}

\bibitem[\protect\citeauthoryear{{S{\'a}nchez}, {Crocce}, {Cabr{\'e}}, {Baugh}
  \& {Gazta{\~n}aga}}{{S{\'a}nchez} et~al.}{2009}]{Sanchez2009}
{S{\'a}nchez} A.~G.,  {Crocce} M.,  {Cabr{\'e}} A.,  {Baugh} C.~M.,
  {Gazta{\~n}aga} E.,  2009, \mn@doi [\mnras]
  {10.1111/j.1365-2966.2009.15572.x}, \href
  {http://adsabs.harvard.edu/abs/2009MNRAS.400.1643S} {400, 1643}

\bibitem[\protect\citeauthoryear{{Sch{\"a}fer}}{{Sch{\"a}fer}}{2009}]{Schafer2009}
{Sch{\"a}fer} B.~M.,  2009, \mn@doi [International Journal of Modern Physics D]
  {10.1142/S0218271809014388}, \href
  {http://adsabs.harvard.edu/abs/2009IJMPD..18..173S} {18, 173}

\bibitem[\protect\citeauthoryear{Sch\"{a}fer \& Koyama}{Sch\"{a}fer \&
  Koyama}{2008}]{Schafer2008}
Sch\"{a}fer B.~M.,  Koyama K.,  2008, Monthly Notices of the Royal Astronomical
  Society, 385, 411

\bibitem[\protect\citeauthoryear{Sch\"{a}fer \& Merkel}{Sch\"{a}fer \&
  Merkel}{2012}]{Schafer2012}
Sch\"{a}fer B.~M.,  Merkel P.~M.,  2012, Monthly Notices of the Royal
  Astronomical Society, 421, 2751

\bibitem[\protect\citeauthoryear{{Sheth} \& {Tormen}}{{Sheth} \&
  {Tormen}}{1999}]{Sheth1999}
{Sheth} R.~K.,  {Tormen} G.,  1999, \mn@doi [\mnras]
  {10.1046/j.1365-8711.1999.02692.x}, \href
  {http://adsabs.harvard.edu/abs/1999MNRAS.308..119S} {308, 119}

\bibitem[\protect\citeauthoryear{{Sheth}, {Mo}  \& {Tormen}}{{Sheth}
  et~al.}{2001}]{Sheth2001}
{Sheth} R.~K.,  {Mo} H.~J.,   {Tormen} G.,  2001, \mn@doi [\mnras]
  {10.1046/j.1365-8711.2001.04006.x}, \href
  {http://adsabs.harvard.edu/abs/2001MNRAS.323....1S} {323, 1}

\bibitem[\protect\citeauthoryear{{Sunyaev} \& {Zeldovich}}{{Sunyaev} \&
  {Zeldovich}}{1980}]{Sunyaev1980b}
{Sunyaev} R.~A.,  {Zeldovich} I.~B.,  1980, \mn@doi [\araa]
  {10.1146/annurev.aa.18.090180.002541}, \href
  {http://adsabs.harvard.edu/abs/1980ARA%26A..18..537S} {18, 537}

\bibitem[\protect\citeauthoryear{{Zel'Dovich}}{{Zel'Dovich}}{1970}]{zeldovich1970}
{Zel'Dovich} Y.~B.,  1970, \aap, \href
  {http://adsabs.harvard.edu/abs/1970A%26A.....5...84Z} {5, 84}

\makeatother
\end{thebibliography}

\label{lastpage}

\end{document}